\documentclass[twocolumn, tighten,]{aastex631}

\shorttitle{An embedded Disk (eDisk) in the IceAge: Investigating the jet and outflow from CED 110 IRS4}
\shortauthors{Narang et al.}
\begin{document}

\title{An embedded Disk (eDisk) in the IceAge: Investigating the jet and outflow from CED 110 IRS4}

\author[0000-0002-0554-1151]{Mayank Narang}
\affiliation{Academia Sinica Institute of Astronomy and Astrophysics, 11F of Astronomy-Mathematics Building, AS/NTU, \\ No.\ 1, Sec.\ 4, Roosevelt Rd, Taipei 106216, Taiwan}

\author[0000-0003-0998-5064]{Nagayoshi Ohashi}
\affiliation{Academia Sinica Institute of Astronomy and Astrophysics, 11F of Astronomy-Mathematics Building, AS/NTU, \\ No.\ 1, Sec.\ 4, Roosevelt Rd, Taipei 106216, Taiwan}

\author[0000-0002-6195-0152]{John J. Tobin}
\affiliation{National Radio Astronomy Observatory, 520 Edgemont Rd, Charlottesville, VA, 22903, USA} 

\author{M. K. McClure}
\affiliation{Leiden Observatory, Leiden University, PO Box 9513, NL–2300 RA Leiden, The Netherlands}

\author[0000-0001-9133-8047]{Jes K. J\o rgensen}
\affiliation{Niels Bohr Institute, University of Copenhagen, \O ster Voldgade 5-7, 1350, Copenhagen, Denmark}

\author[0000-0003-4361-5577]{Jinshi Sai (Insa Choi)}
\affiliation{Academia Sinica Institute of Astronomy and Astrophysics, 11F of Astronomy-Mathematics Building, AS/NTU, \\ No.\ 1, Sec.\ 4, Roosevelt Rd, Taipei 106216, Taiwan}

\author{eDisk + IceAge Team.}

\begin{abstract}
We present a comprehensive study of the large-scale structure, jet and outflow morphology, and kinematics of the Class 0/I protostellar binary Ced 110 IRS4, using JWST NIRCam (F150W and F410M) and MIRI MRS observations from the JWST ERC program IceAge, along with ALMA data from the Early Planet Formation in Embedded Disks (eDisk) program. NIRCam images, combined with ALMA continuum and CO data, reveal arc-like structures ($\sim$1100 au), suggesting a dense envelope around the protostars. We detect disk shadows from both protostars in F150W. The MIRI MRS IFU data reveal a jet from both protostars in multiple [Fe II] lines,  [Ar II] 6.99 $\mu$m and [Ne II] 12.81 $\mu$m, marking the first detection of a jet from the system. {The [Fe II] (5.34 $\mu$m) jet from Ced 110 IRS4A has a width of $\leq$ 51~au at the protostellar location, with a large opening angle of 23\arcdeg{} $\pm$ 4\arcdeg.} After inclination correction, the jet velocity is 124 km s$^{-1}$, corresponding to a dynamical timescale of 25 years. {The molecular H$_2$ outflow displays a distinct morphology resembling two hemispheres placed back-to-back.} The consistent H$_2$ emission extent across transitions, differing from previous observations of protostellar outflows detected with JWST, suggests that MHD disk winds may not drive the observed outflow. {We find that the upper limit to the width of the outflow at the protostellar location is 130 $\pm$ 10 au which is smaller than the disk diameter of 183.4 {$\pm$ 0.4 au}  but much larger than width of the [Fe II] jet.}
\end{abstract}

\section{Introduction}

Class~0 protostars represent the earliest phase of star formation and are empirically defined as having bolometric temperature T$_\mathrm{bol}<70$ K or L$_\mathrm{smm}$/L$_\mathrm{bol}$ $>$0.005 \citep[e.g.,][]{1993ApJ...406..122A,2007prpl.conf..117W,2009ApJS..181..321E,2014prpl.conf..195D,2024arXiv240315550T}. Class 0 protostars possess dense circumstellar envelopes that feed the protostellar disk  \citep{U76,C81}, which in turn accretes onto the central star. These protostars also drive strong bipolar jets/outflows   \citep{2007prpl.conf..261S, 2007prpl.conf..277P, 2016ARA&A..54..135H, 2014prpl.conf..451F, 2016ARA&A..54..491B, 2021NewAR..9301615R} which remove angular momentum from protostellar disks and thus {maintain} the accretion.

The low mass accretion rates measured from more evolved protostars Class I and Flat-spectrum sources (using optical and near-IR spectra) suggest that most of the accretion occurs in the earlier Class 0 phase \citep{2004ApJ...616..998W, F22, F23, M23}. However, protostars in their primary accretion phase are deeply embedded within dense envelopes, {making it difficult for direct detection of accretion tracers in the visible and near-IR wavelengths.} The protostellar jets/outflows however can escape their dense envelopes making it possible to study the outflow rates at optical and near-IR wavelengths \citep[e.g.,][]{2002AJ....123..362R,2013MNRAS.433.2226R}. Accretion-driven outflows and jets also play vital roles in shaping the Initial Mass Function (IMF) \citep[e.g.,][]{2022MNRAS.515.4929G,2024A&A...683A..13L}. They can inject energy and momentum into the surroundings and disperse a significant fraction of the protostellar envelope, which limits star formation efficiency \citep[e.g.,][]{2010ApJ...710L.142F,2014ApJ...784...61O,2014prpl.conf..451F,2015MNRAS.450.4035F,2021ApJ...911..153H,2023ApJ...954...93A}.

Class~0 protostars have been extensively studied at (sub)mm wavelengths \citep[e.g.,][]{1993ApJ...406..122A,2005ApJ...633L.129B,2007ApJ...659..479J,2008ApJS..175..277D,2009A&A...507..861J,2019ApJ...871..221M, Lee21, 2021A&A...648A..45P} and NIR/MIR \citep[e.g.,][]{2002AJ....123..362R,2013MNRAS.433.2226R,2016ApJ...828...52W,2022ApJ...941L..13Y, 2023ApJ...951L..32H,2024A&A...685A..27B, 2024A&A...686A..71N, 2024A&A...688A..29S,2024arXiv241006697T,2024arXiv240916061C}. {These observations have revealed the presence of highly collimated, fast-moving molecular jets and wide-angled, slow-moving molecular outflows originating from the vicinity of protostars, although molecular jets may not be universally detected toward all protostars \citep[e.g.,][]{Lee17,2020A&ARv..28....1L, 2021NewAR..9301615R, 2023arXiv230615346D}.} Using isotopologues of CO and other dense gas tracers,  Atacama Large Millimeter/submillimeter Array (ALMA) has detected the cool gas that dominates the mass of the outflow and is too cold to be detected by JWST. JWST has detected and mapped warm gas in accretion flows and outflow shocks \citep{Ray23,2023arXiv231207807R,2024arXiv240613084G, 2024arXiv240916061C} with an angular resolution comparable to ALMA. JWST has also detected the atomic jet \citep{2022ApJ...941L..13Y, 2023ApJ...951L..32H, 2023A&A...673A.121B,2023arXiv231003803F,2023arXiv231014061N} and direct signatures of accretion such as the H-I lines \citep{2023ApJ...951L..32H,2023arXiv231003803F} and OH emission \citep{2024arXiv240407299N}.

\begin{figure*}
\href{}{}\centering
\includegraphics[width=0.5\linewidth]{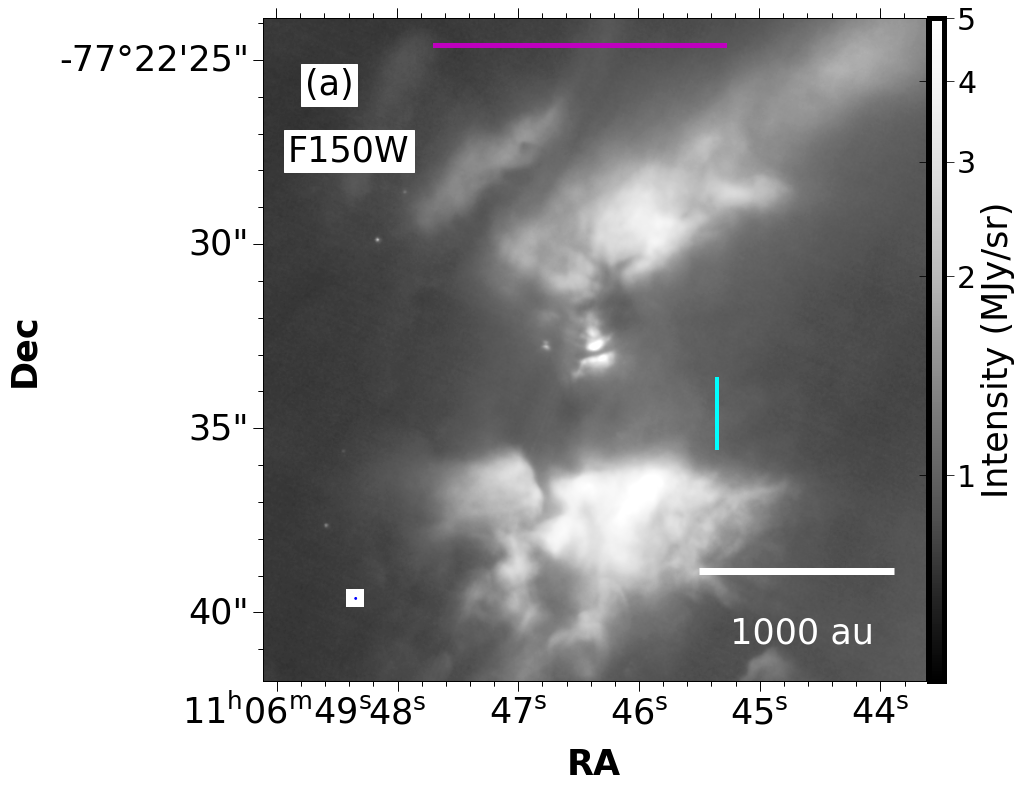}\includegraphics[width=0.5\linewidth]{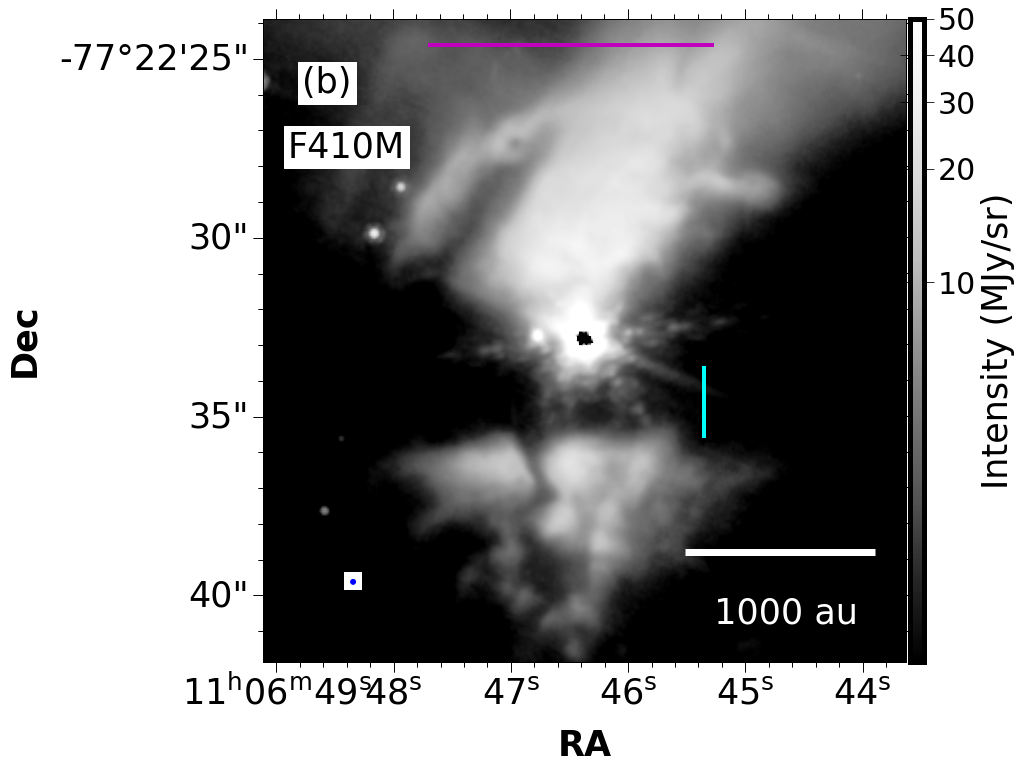}
\includegraphics[width=0.5\linewidth]{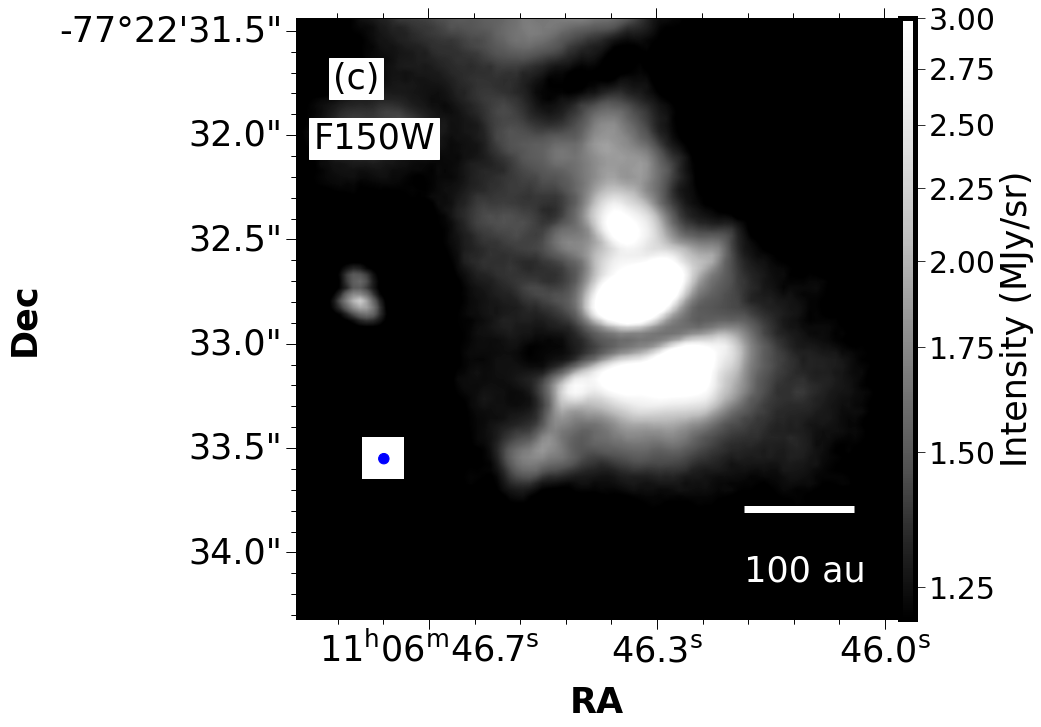}\includegraphics[width=0.5\linewidth]{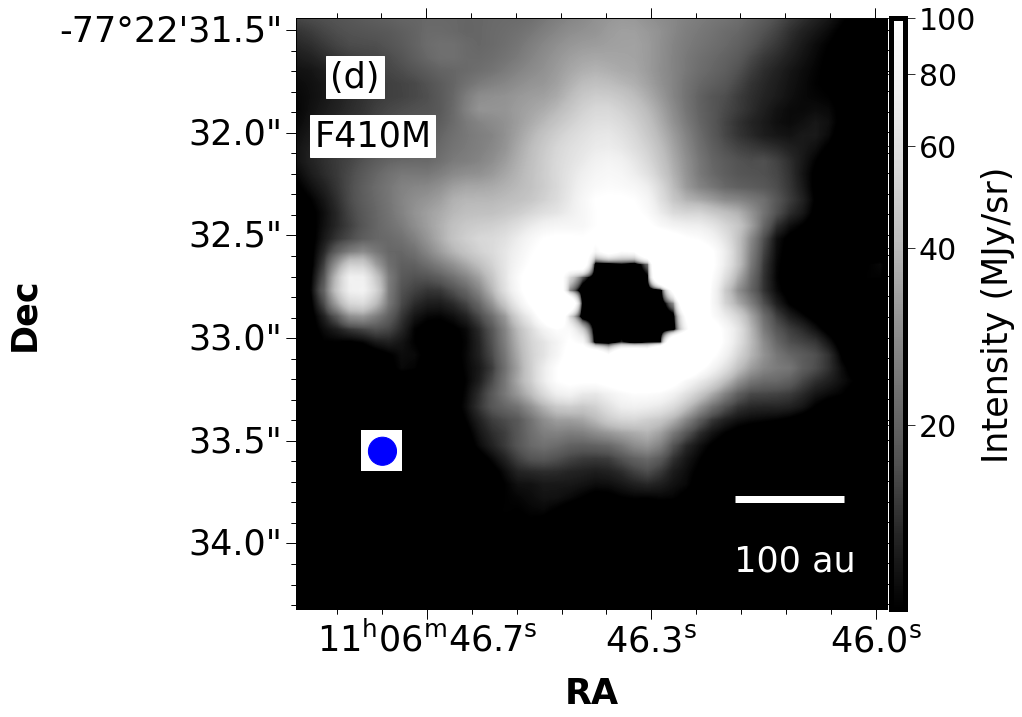}
\caption{The NIRCam F150W and F410M filter image of the Ced 110 IRS4 region in Greyscale. The NIRCam FWHM is shown in the bottom left corner as blue. The scale bar in white is also shown in the bottom right corner. The beam sizes are given in Table \ref{NIRcam}. {The arc-like structure is marked with magenta line and the dark band is marked with cyan line.} The same intensity scaling is followed in Figure \ref{Fig1} for the NIRCam observations.  }
\label{Fig0}
\end{figure*}

Cederblad (Ced) 110 IRS4 (eHOPS-cha-3) is a Class 0/I protostellar system (L$_\mathrm{bol}$ = 1.73~L$_\odot$, T$_\mathrm{bol}$ = 74~K; \citealt{P23,2023ipac.data.I553P} \footnote{irsa.ipac.caltech.edu/data/Herschel/eHOPS/overview.html}; also see \citealt{2023ApJ...951....8O})  in the Chamaeleon (Cha) I dark cloud {\citep{1991MNRAS.251..303P,2007ApJ...664..964H,2009A&A...508..259V,2011ApJS..193...11M}} located at a distance of about 189 pc from Earth. Ced 110 IRS4, is the brightest object in the  Ced 110 region \citep{2001A&A...376..907P} and is associated with near-IR bipolar nebulosity \citep{1999A&A...352L..73Z,2001A&A...376..907P,2005A&A...435..595P}.  High-resolution ALMA observations of the Ced 110 IRS~4 have shown that the system is a binary consisting of Ced 110 IRS~4A (RA = 11:06:46.369; Dec =  -77:22:32.88) and Ced 110 IRS~4B (RA = 11:06:46.772; Dec = -77:22:32.76) \citep{2023ApJ...951....8O,2023ApJ...954...67S} with a projected separation of $\sim$ 1\arcsec.3 or 250 au. \cite{2023ApJ...954...67S} found that Ced 110 IRS~4A has a stellar mass of 1.21~--~1.45~M$_\odot$  while Ced 110 IRS~4B  is a  substellar companion with a mass of 0.02-0.05 $M_\odot$. {They estimated the dust disk radii for Ced 110 IRS4A and IRS4B to be $91.7 \pm 0.2$~au  and $ 33.6 \pm 0.6$~au respectively. The disk masses range between 38.85 - 97.43 M$_\oplus$ and 0.825 - 2.07 M$_\oplus$ for Ced 110 IRS4A and Ced 110 IRS4B. \cite{2023ApJ...954...67S}  also suggested that the northern side of the disk might be the near-side. The physical properties of the disks derived from the {ALMA eDisk} data are important for us to understand the outflow/jet observed with JWST. }
\begin{figure*}
\href{}{}\centering
\includegraphics[width=0.5\linewidth]{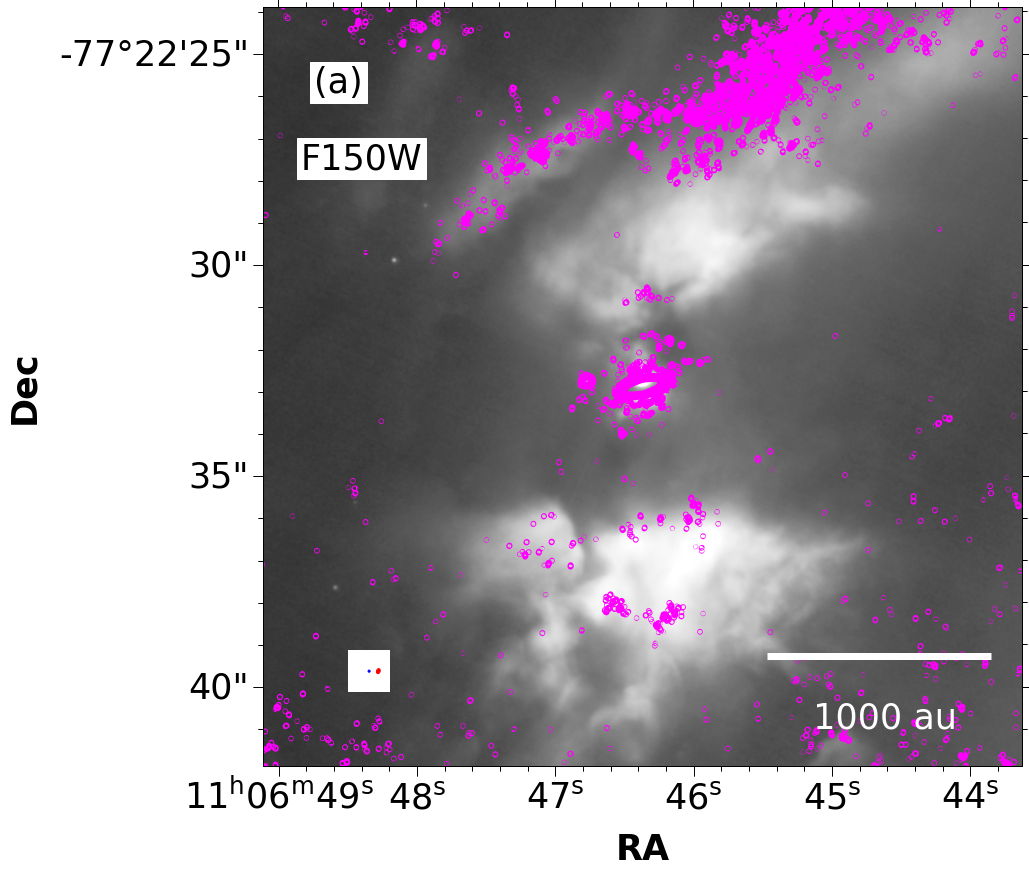}\includegraphics[width=0.5\linewidth]{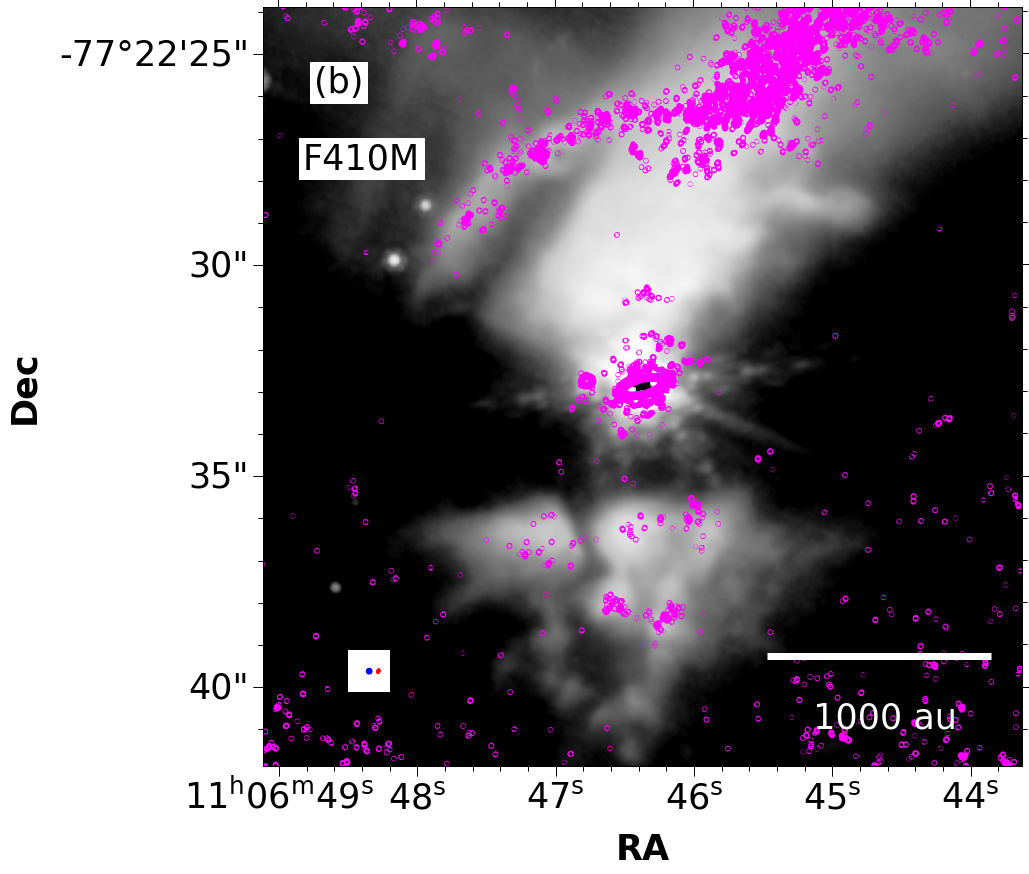}
\includegraphics[width=0.5\linewidth]{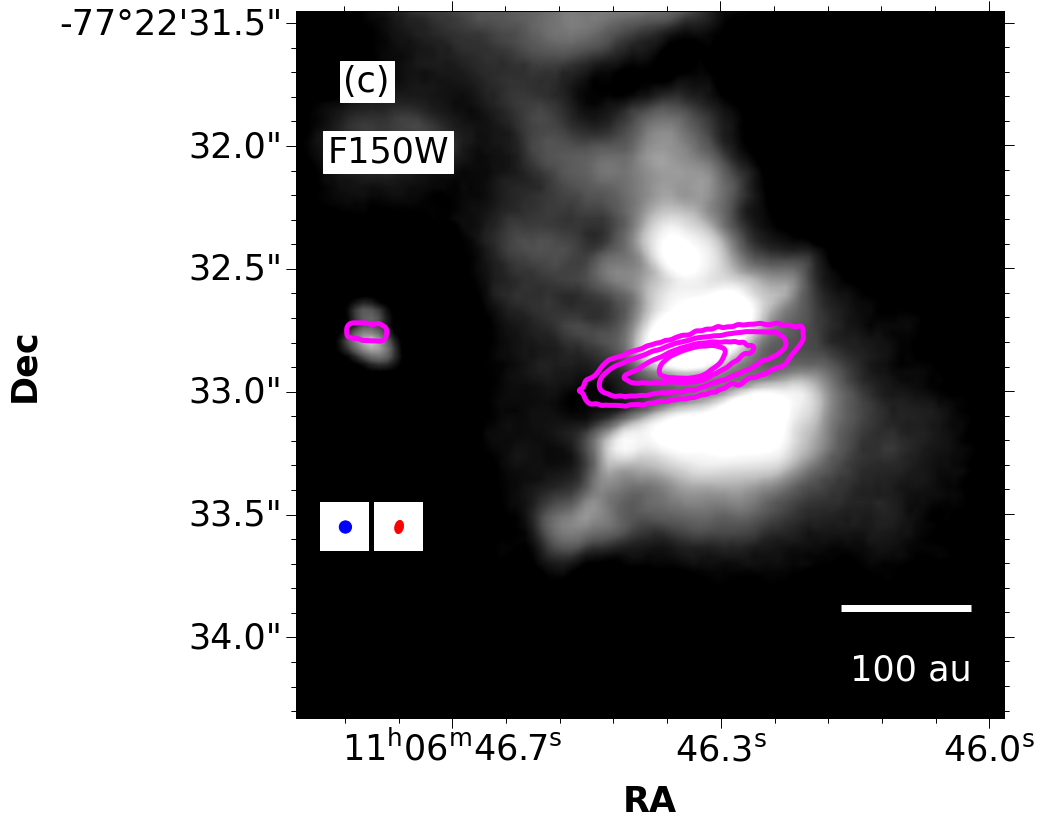}\includegraphics[width=0.5\linewidth]{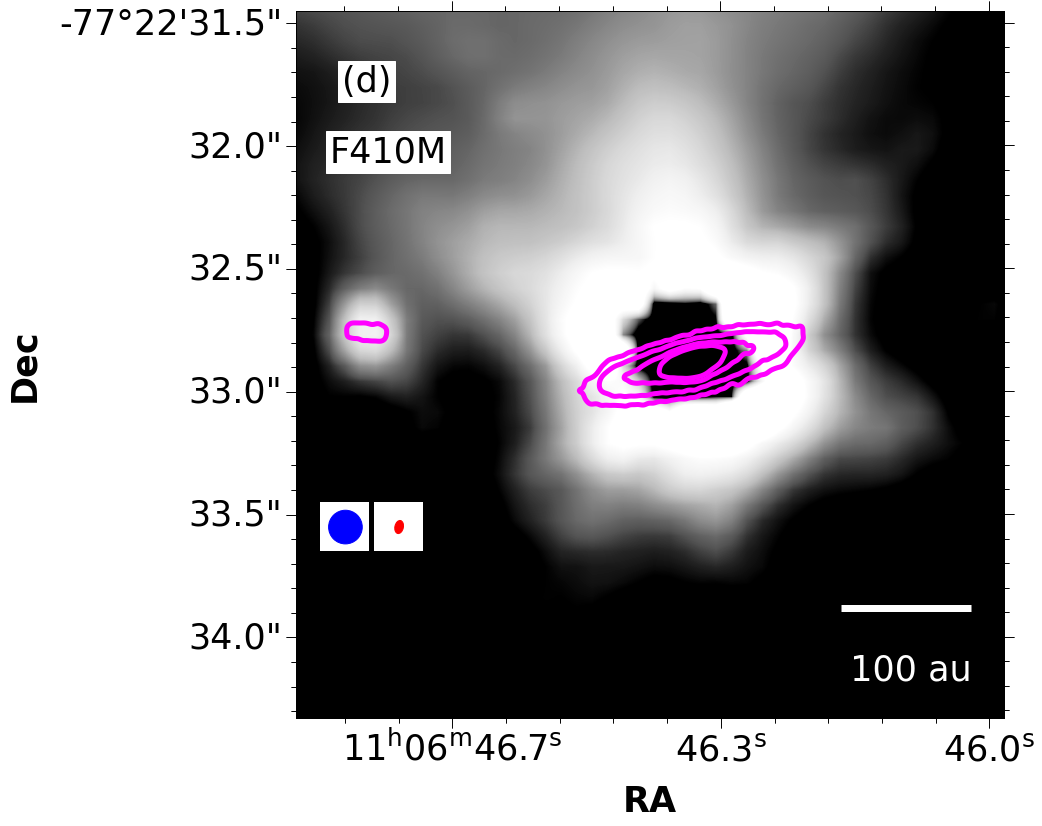}
\caption{The NIRCam F150W and F410M filter image of the Ced 110 IRS4 region in Greyscale is shown with the ALMA 1.3 mm continuum map overlaid as magenta contours. The 1.3 mm continuum maps, with robust parameters of 2.0 and 0.0, are overlaid on the wide and zoomed-in views of the NIRCam image, respectively. The contours are  3, 10, 30, 70, and 100 $\times$~$\sigma$, with $\sigma_{robust = 2} \,=14\, \mu$Jy and $\sigma_{robust =0} \,=16\, \mu$Jy with 10, 30, 70, and 100 $\times$~$\sigma$. The NIRCam FWHM and the ALMA beams are shown in the bottom left corner as blue and red ellipses respectively.  The scale bar in white is also shown in the bottom right corner. The beam sizes and PA are given in Table \ref{Table1} and Table \ref{NIRcam}. }
\label{Fig1}
\end{figure*}

Arc-like structures on the northern side of the Ced 110 IRS4 at a radius of $\sim$1100 au ($\sim$6$''$) were detected in the NIR by \citet{1999A&A...352L..73Z,2001A&A...376..907P, 2005A&A...435..595P}. The higher resolution ALMA observations of several species, including  C$^{18}$O $J=2-1$,  SO $J_N=6_5$~-~$5_4$, H$_2$CO, and c-C$_3$H$_2$ found similar structure and were able to study their kinematics and substructures with the high spatial and spectral resolution offered by the ALMA data \citep{2023ApJ...954...67S}.

The NIR/MIR emission from Ced 110 IRS4 system has been studied  with Spitzer \citep{2010A&A...519A...3L,2011ApJS..193...11M}. These studies not only detected the gas phase {emission from} [Fe II], [Ne II] and molecular H$_2$, but also detected strong absorption features at 6.0 \micron{} due to H$_2$O ice, at 6.8 \micron{} due to CH$_3$OH, NH$^+_4$ , and strongly polar H$_2$O ice, and at 15.2 \micron{} due to CO$_2$ ice along with the 10 \micron{} silicate feature. {These observations however lacked the resolution as well as the sensitivity to spatially map these features.} 

Recently \cite{2024arXiv241119651R} carried out a detailed analysis of the ice features seen towards Ced 110 IRS4A and IRS4B using JWST Near Infrared Camera (NIRCam)  Wide-Field Slitless Spectroscopy (WFSS),  Near InfraRed Spectrograph (NIRSpec) and Mid-Infrared Instrument (MIRI) Medium Resolution Spectroscopy (MRS).  They found that the ice abundances for Ced 110 IRS4A and IRS4B was similar to the starless regions in the same molecular cloud and both protostars showed hints of thermal processing  of ices. This study however did not analyze the gas phase emission from the system.

Despite multiple submillimeter observations of Ced 110 IRS4 \citep{2007ApJ...664..964H, 2009A&A...508..259V, 2023ApJ...954...67S}, no definitive signatures of outflows or jets have been detected from the system. Moreover, the rich gas-phase spectra observed towards the two protostars \citep{2024arXiv241119651R} necessitate further investigation. To better understand the structure revealed by ALMA submillimeter observations (both continuum and molecular line emission), the scattered light morphology, and to resolve whether Ced 110 IRS4 is associated with a jet or a wide-angle outflow, we conducted a joint analysis using ALMA and JWST observations.

Our study leverages high-resolution imaging from JWST's Near Infrared Camera (NIRCam) \citep{2005SPIE.5904...30H} and Medium Resolution Spectroscopy (MRS) IFU data from the Mid-Infrared Instrument (MIRI) \citep{2015PASP..127..584R, 2015PASP..127..595W}. {These datasets \citep{McClure,2024arXiv241119651R} enable a systematic} examination of near-infrared (NIR) and mid-infrared (MIR) jet and outflow kinematics and morphology in the Ced 110 IRS4 region, traced through atomic and fine-structure lines as well as molecular H$_2$. A detailed analysis of the mass accretion and outflow rates will be presented in Narang et al. (in preparation). In Section 2, we describe the JWST and ALMA observations and the data reduction process. Section 3 presents our analysis of the large-scale outflow structures and the morphology of the inner $\sim$1000 au surrounding the protostar. Section 4 discusses the implications of our findings, and Section 5 summarizes the study's conclusions.

\begin{figure*}
\centering
\includegraphics[width=0.5\linewidth]{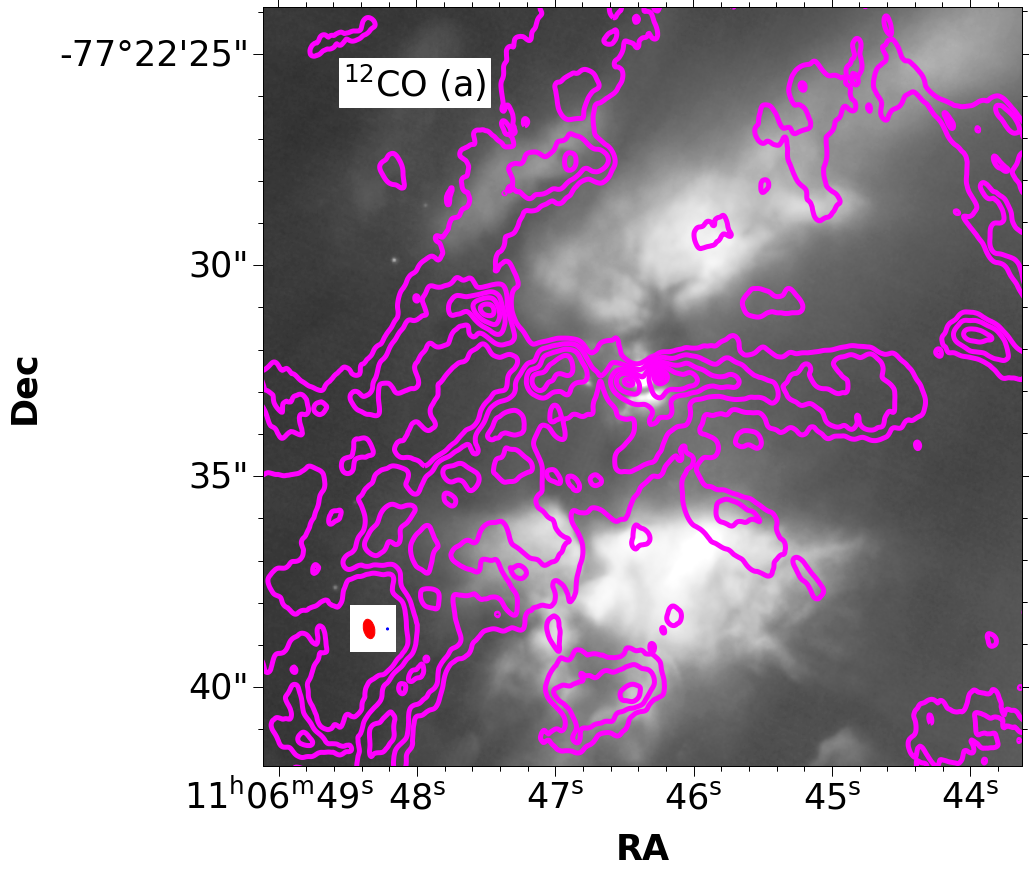}\includegraphics[width=0.5\linewidth]{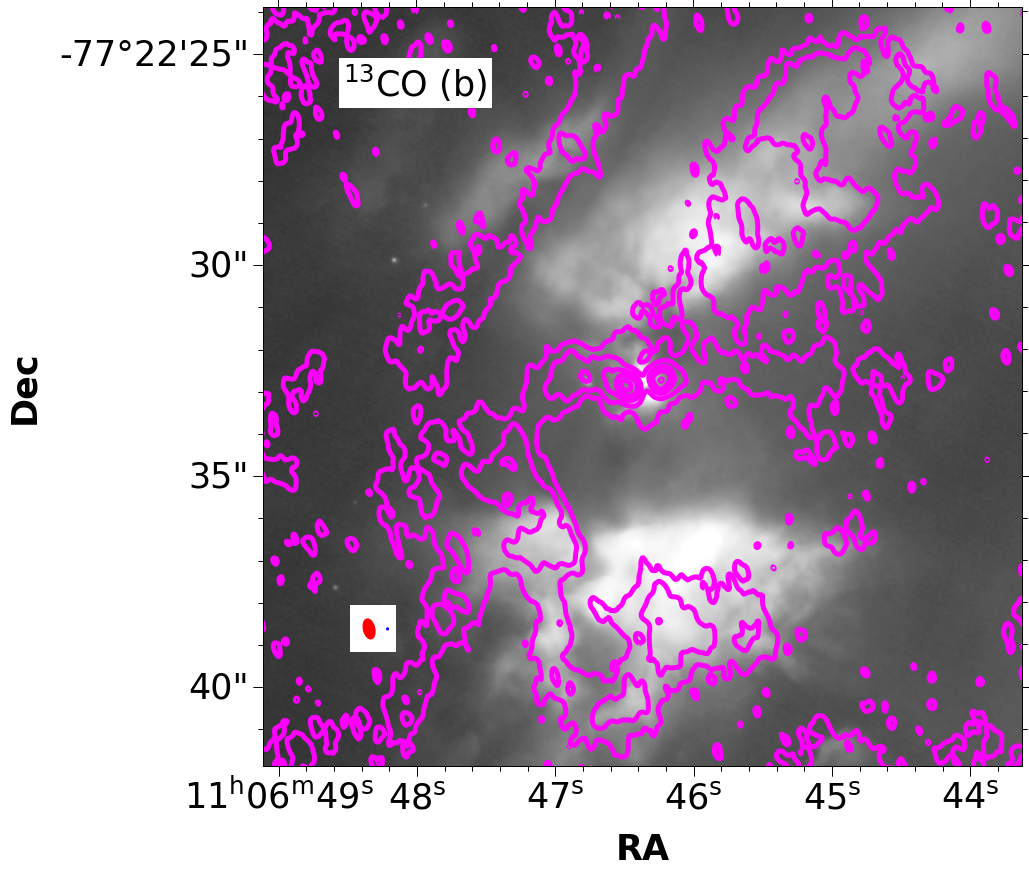}
\includegraphics[width=0.58\linewidth]{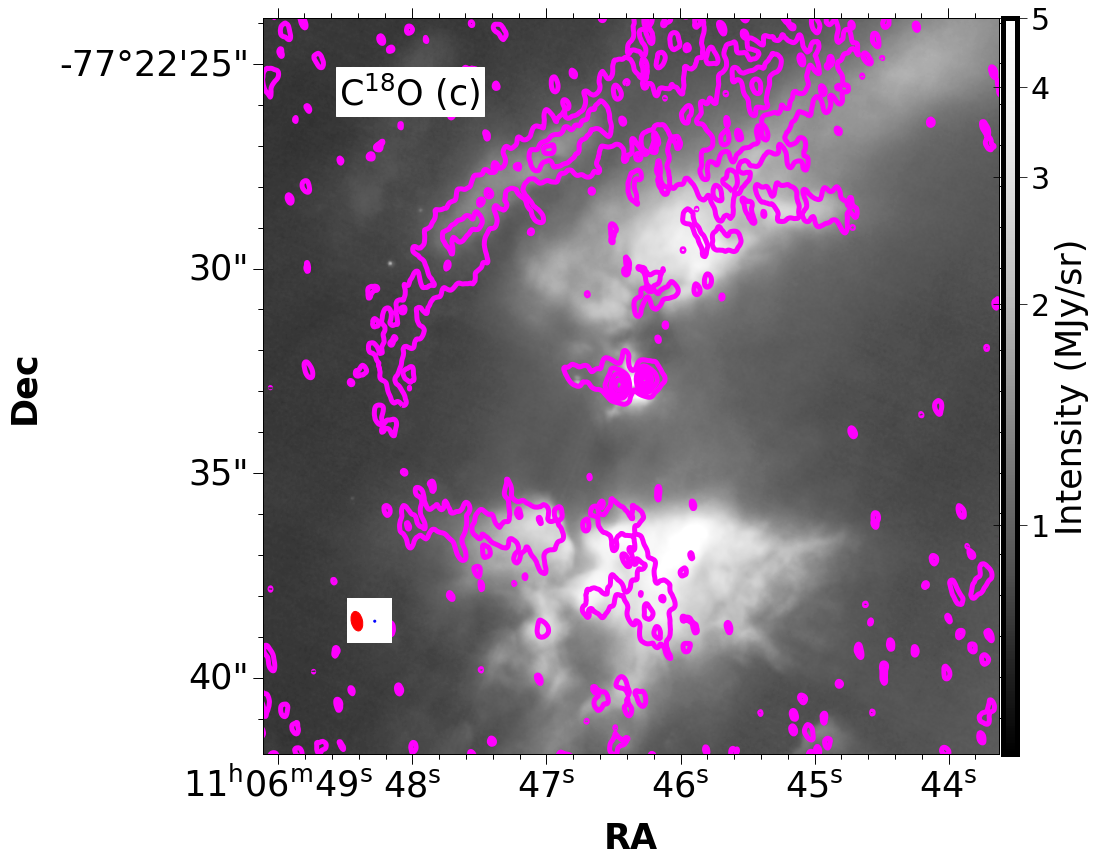}
\caption{The NIRCam F150W filter image of the Ced 110 IRS4 region with the ALMA (SB) CO isotopologues integrated intensity image overlaid as magenta contours. The contours for  $^{12}$CO and  $^{13}$CO are 15\%, 30\%, 50\%, 70\%, 90\%, and 99\% $\times F_{max}$, where $F_{max}$ for $^{12}$CO is 0.36 Jy/beam km.s$^{-1}$; $F_{max}$ for $^{13}$CO is 0.17 Jy/beam km.s$^{-1}$. The contours for C$^{18}$O are  30\%, 50\%, 70\%, 90\%, and 99\% $\times F_{max}$,  where $F_{max}$ for C$^{18}$O is 0.07 Jy/beam km.s$^{-1}$. The NIRCam FWHM and the ALMA beams are shown in the bottom left corner as blue and red ellipses respectively. The beam sizes and PA are given in Table \ref{Table1} and Table \ref{NIRcam}. All NIRCam images have the same intensity scale. }
\label{Fig2a}
\end{figure*}

\begin{figure*}
\centering
\includegraphics[width=0.5\linewidth]{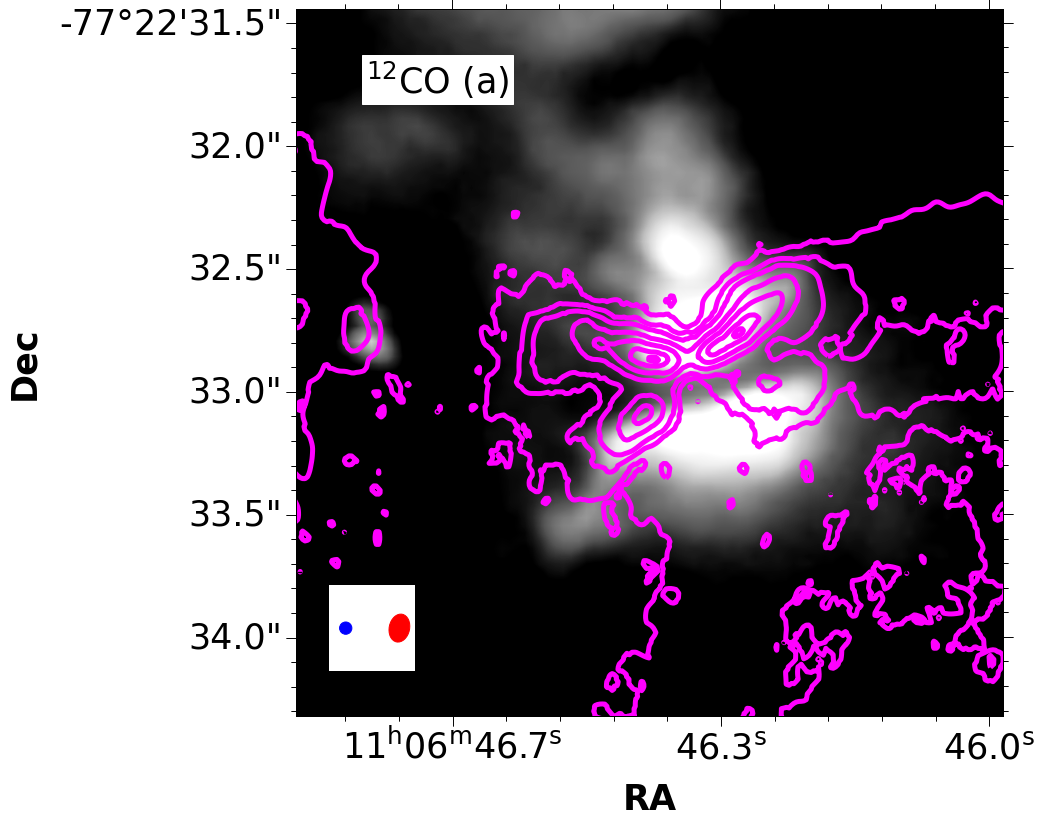}\includegraphics[width=0.5\linewidth]{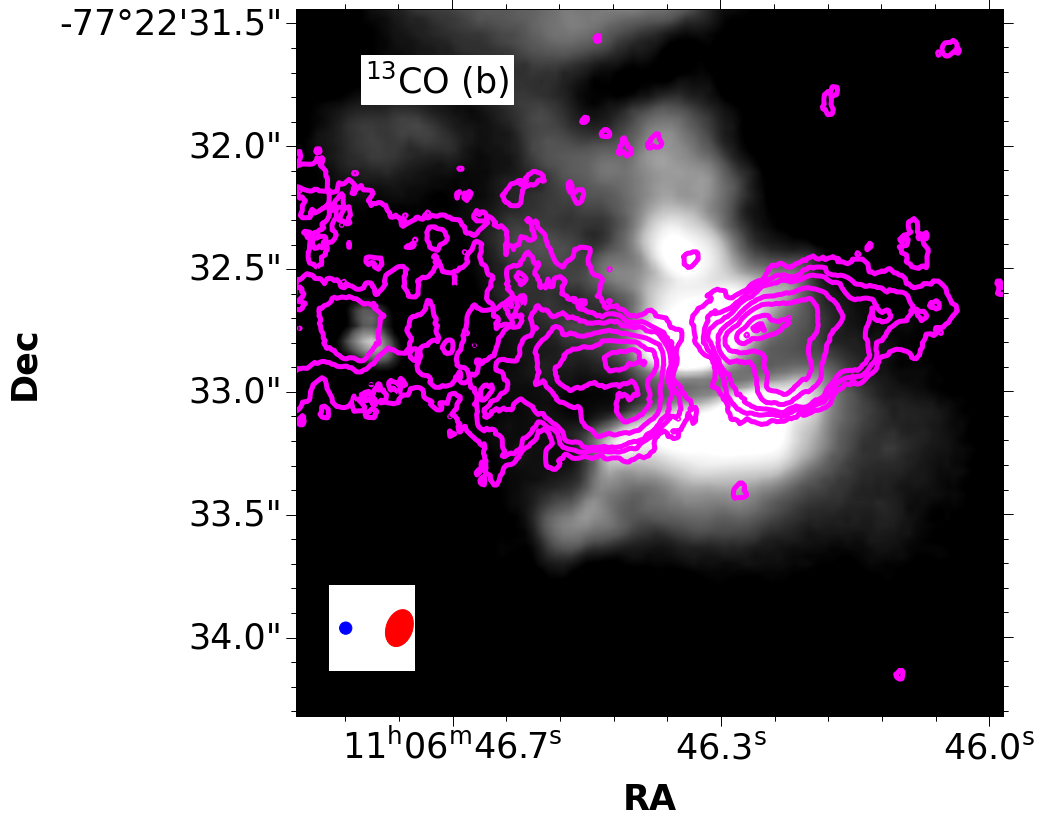}
\includegraphics[width=0.58\linewidth]{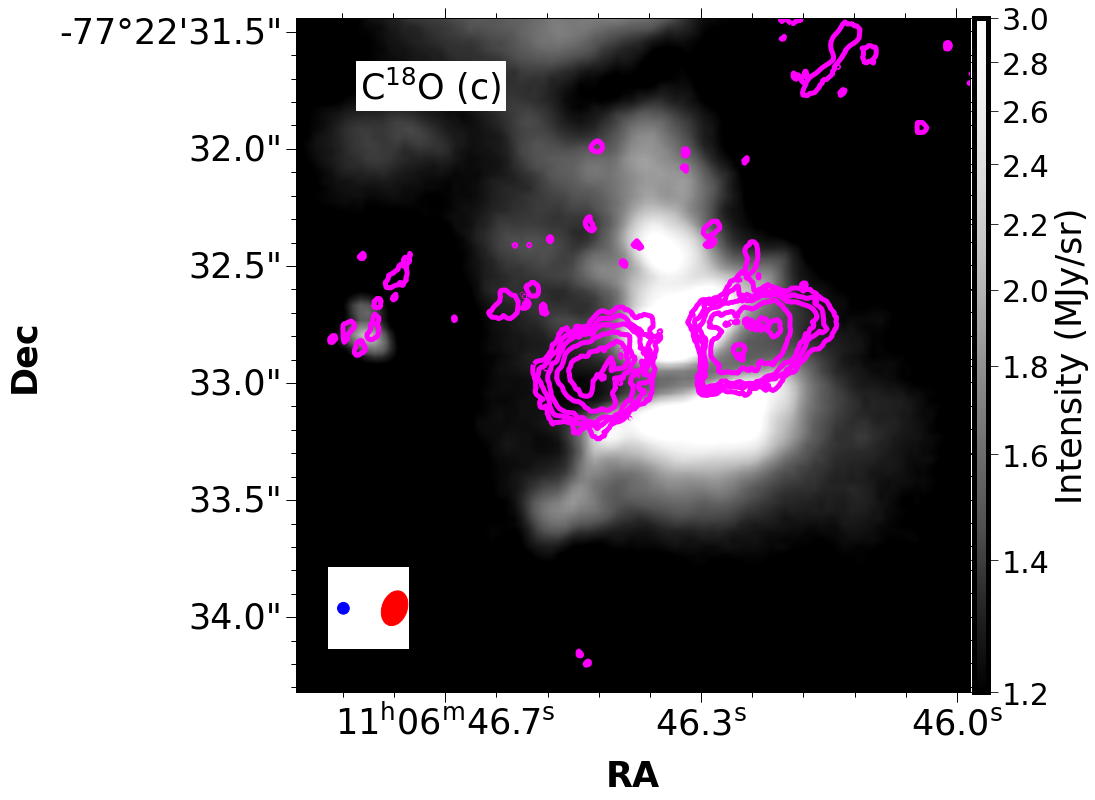}
\caption{The NIRCam F150W filter image of the Ced 110 IRS4 region with the ALMA (SBLB) CO isotopologues integrated intensity image overlaid as magenta contours. The contours for are 15\%, 30\%, 50\%, 70\%, 90\%, and 99\% $\times F_{max}$, where $F_{max}$ for $^{12}$CO is 150 mJy/beam km.s$^{-1}$; $F_{max}$ for $^{13}$CO is 60 mJy/beam km.s$^{-1}$ and $F_{max}$ for C$^{18}$O is 30 mJy/beam km.s$^{-1}$. The NIRCam FWHM and the ALMA beams are shown in the bottom left corner as blue and red ellipses respectively. The beam sizes and PA are given in Table \ref{Table1} and Table \ref{NIRcam}. All NIRCam images have the same intensity scale. }
\label{Fig4}
\end{figure*}

\section{Observations and Data reduction}

\subsection{JWST observations}

The NIRCam and MIRI MRS observations of  Ced 110 IRS~4 were carried out as part of the JWST Director’s Discretionary Early Release Science Program IceAge: Chemical Evolution of Ices during Star Formation (Program ID 1309; PI Melissa McClure, {Adwin Boogert, Harold Linnartz - in memoriam); \citet{McClure,2024arXiv241119651R}}. The NIRCam imaging was carried out with the  F140M,  F150W,  and F410M filters. The NIRCam reduction was carried out using the default reduction script. The data were reduced using the pipeline version {1.14.0} and the JWST Calibration References Data System context version {jwst\_1242.pmap}. The NIRCam images have an empirical FWHM of 0\farcs046, 0\farcs049 and 0\farcs133 for the  F140M,  F150W,  and F410M filters respectively\footnote{https://jwst-docs.stsci.edu/jwst-near-infrared-camera\/nircam-performance\/nircam-point-spread-functions} .

\begin{figure*}
\href{}{}\centering
\includegraphics[width=0.5\linewidth]{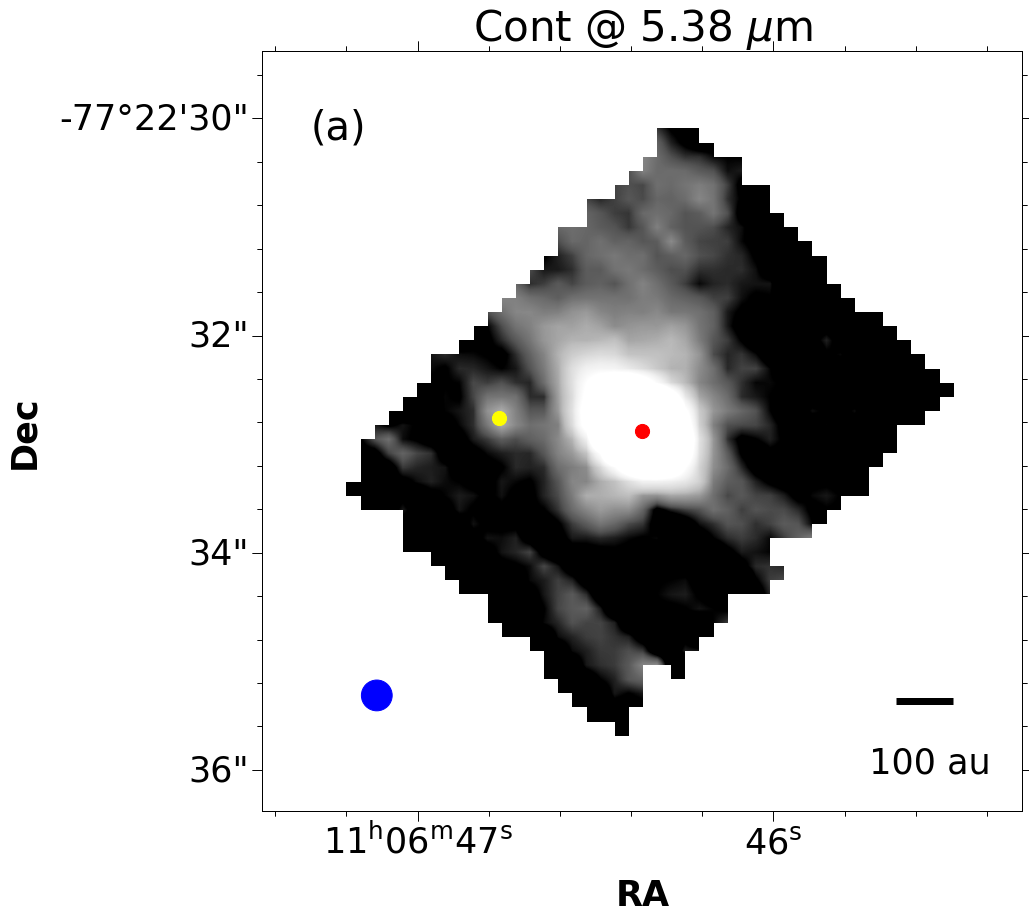}\includegraphics[width=0.5\linewidth]{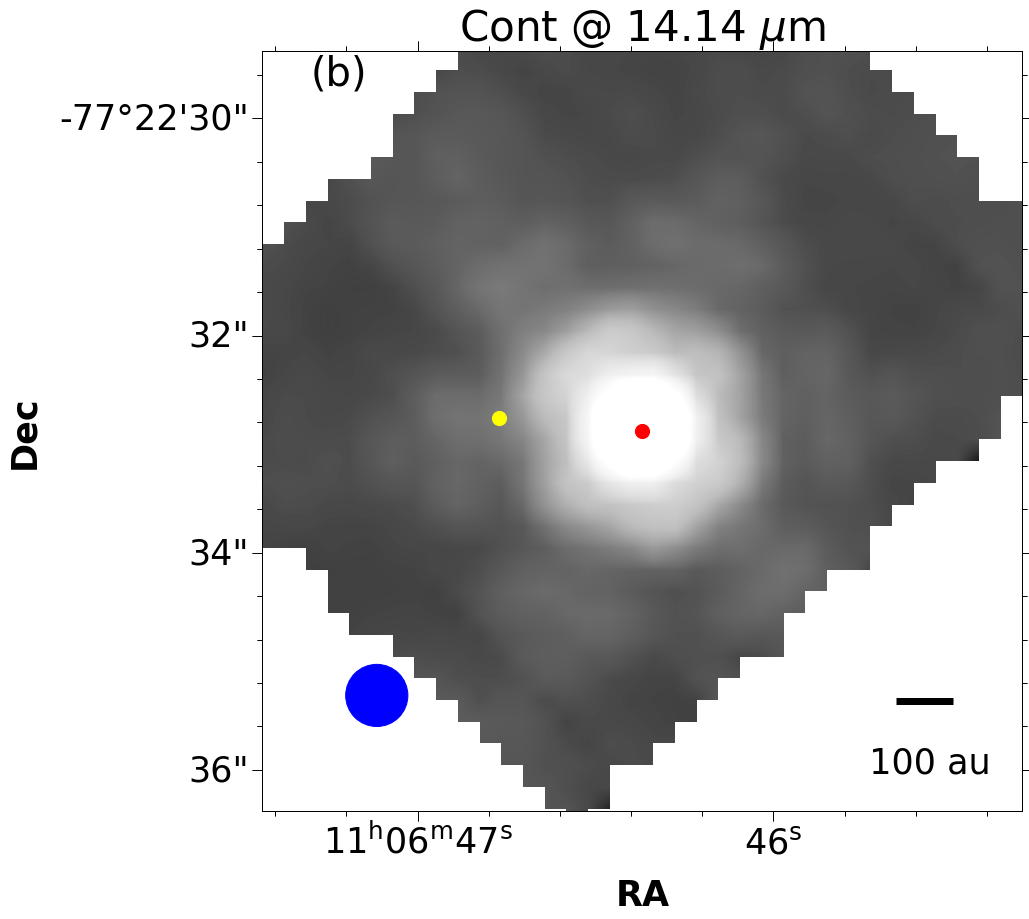}
\caption{The MIRI MRS continuum image of the Ced 110 IRS4 region in Greyscale. The ALMA 1.3 mm position for Ced 110 IRS4A is shown as a red circle and for IRS4B is shown as yellow circle.   The scale bar corresponding to 100 au is shown in the bottom right corner. The MIRI FWHM based on \cite{2023AJ....166...45L} is shown in the bottom left corner as a blue circle. }
\label{MIRI_cont}
\end{figure*}

The MIRI MRS observations of Ced 110 IRS~4 were carried out on  2023 July 12, almost a year after the NIRCam observations. Ced 110 IRS4 was observed using all four channels of MIRI MRS. Each channel was exposed for 1665 sec, using the ``FASTR1" readout pattern with 150 Groups/Int and with a single integration. The data were then reduced using the JWST pipeline, version {1.14.1.dev49+g9e857a52}, and the calibration file CRDS {jwst\_1215.pmap}. The MIRI data was reduced using the standard {Stage}-1 JWST pipeline \textit{Detector1Pipeline} to reduce the MIRI MRS data starting from \textit{uncal} data which was followed by  {Stage}-2 JWST pipeline \textit{Detector2Pipeline}. The {Stage}-3 \textit{Spec3Pipeline} step was ran with the \textit{CubeBuildStep}  set to \textit{'band'} mode such that each channel and each band are reduced as a separate fits file. {Although most cosmic ray events are flagged and rejected during the {Stage}-1 process while constructing the uncalibrated slope image, some outliers still manage to pass through both the {Stage}-1 and {Stage}-2\footnote{https://jwst-docs.stsci.edu/jwst-science-calibration-pipeline/{Stage}s-of-jwst-data-processing/calwebb\_spec3\#gsc.tab=0}. Therefore we also turned on the outlier detector in {Stage}-3, which helped remove many of the artifacts from the spectral cube.} 

\subsection{ALMA observations}

The ALMA observations of Ced110 IRS4  were carried out during  ALMA's Cycle 7 and 8 at ALMA Band 6 with antenna configurations of C-5 and C-8, as a part of the ALMA large program Early Planet Formation in Embedded Disks (eDisk) (2019.1.00261.L: PI N.\ Ohashi \citealt{2023ApJ...951....8O}). {An overview of the data and first results about Ced 110 IRS4 from the ALMA  eDisk observations were presented by  \cite{2023ApJ...951....8O} and  \cite{2023ApJ...954...67S}.  Thus, we will only present a brief overview of the observational setup and data reduction.} CO isotopologues, SO, and other molecular lines were observed using Frequency Division Mode (FDM). The 1.3 mm (225 GHz) continuum emission was extracted from the line-free channels of the spectral windows, with a maximum window width of 1.875 GHz. The ALMA data was reduced with the Common Astronomy Software Applications package CASA 6.2.1 \citep{2007ASPC..376..127M}. We made Short-Baseline + Long-Baseline (SBLB) continuum images and both Short-Baseline (SB) as well as SBLB images for the line emission from  Ced 110 IRS4 region for varying robust values. The details of the imaging procedure can be found in \cite{2023ApJ...951....8O} and \cite{2023ApJ...954...67S}. We have listed the parameters (robust value, beam size, and beam position angle) for the various maps used in this work in Table \ref{Table1}.

Since we are comparing data from ALMA and JWST in this work, we assessed the offset between ALMA and JWST images by measuring the position of Ced 110 IRS4B. For the NIRCam observations we used the F410 filter. Due to the disk shadow present in the F150 filter{ (see Figure \ref{Fig0} and Figure \ref{Fig1}),} it was more challenging to accurately measure the position in that band. We did not measure the position of IRS4A as it was saturated.  The offset between the NIRCam image and ALMA was found to be approximately 0\farcs055, which is less than one pixel {(pixel-scale for F410M 0\farcs063/pixel)}, indicating an excellent positional match between NIRCam and ALMA observations.  

The MIRI MRS channel-1 observations of Ced 110 IRS4B have contamination from the MIRI MRS PSF of Ced 110 IRS4A and beyond channel 8 $\mu$m the emission from Ced 110 IRS4B drops. Furthermore at shorter wavelengths the contribution from scatter ligh emission is also high. Therefore we have opted to use the 14.14 $\micron$ emission from Ced 110 IRS4A to investigate the positional offset between the JWST MIRI and ALMA position. The MIRI 14.14 $\mu$m continuum position of Ced 110 IRS4A differed from the ALMA 1.3 mm continuum position by 0\farcs2 {(the pixel-scale for MIRI 14.14 $\mu$m is 0\farcs2/pixel)}. To address this, we corrected the astrometry of the MIRI MRS observations by shifting the 14.14 $\mu$m continuum position of the protostar to align with the ALMA 1.3 mm continuum position.

\section{Results}
\subsection{Continuum emission with NIRCam \& MIRI MRS}

{Previous NIR observations of Ced 110 IRS4 \citep{2005A&A...435..595P} have shown the existence of a bipolar structure separated by a dark band and a bright central source.} However, those observations were limited to a resolution of 0\farcs7, much larger than the {ALMA eDisk} (see Table \ref{Table1}) or the NIRCam IceAge observations.  To explore the large-scale structure around the Ced 110 IRS4 system, we analyzed the NIRCam F150W and F410M observations alongside ALMA continuum and CO isotopologue data.

Figure \ref{Fig0} shows the NIRCam observations of the Ced 110 IRS4 region in F150W and F410M filters. In Figure \ref{Fig0}(c) and Figure \ref{Fig0}(d), we present a close-up view of the Ced 110 IRS4 region, focusing on the inner $\sim$ 3\arcsec{} region. The NIRCam observations detect not only Ced 110 IRS4A but also Ced 110 IRS4B. The NIRCam images (at large scales; Figure \ref{Fig0}a and Figure \ref{Fig0}b) show bipolar nebulosity with lobes and arc-like structures separated by a dark band that is $\sim$2\arcsec.5 wide. This arc-like structure observed in NIRCam is also detected in the 1.3 mm continuum map (Figure \ref{Fig1}a and Figure \ref{Fig1}b). We also see a bow shock towards the south of the protostars, associated with a craved gap in the nebulosity. 

The zoomed in JWST NIRCam F150W (Figure \ref{Fig0}c) shows a dark lane passing through both Ced 110 IRS4A and IRS4B. The dark lane represents the disk shadow from the nearly edge-on disk around each protostar.  This also indicates that Ced 110 IRS4B has a near edge-on inclination, similar to Ced 110 IRS4A; consistent with what was found by \cite{2023ApJ...954...67S}. We further find that the NIRCam F150W emission is also extended towards the north in the direction of the jet (see section 3.2 and Figure \ref{jet_NIR}). Within the wavelength range of the F150W filter, there are several transitions of [Fe II] lines that have been detected with JWST. {These [Fe II] lines trace the atomic jet from the protostar \citep[e.g.,][]{2024arXiv240418334A} which can be used to determine the outflow rate from the jet. The emission from Ced 110 IRS4A is however saturated in F410M (Figure \ref{Fig0}d) and thus making it difficult to analyze. }

To investigate the relationship between the disk shadow and the dust disk at 1.3 mm, we overlaid the ALMA continuum image onto the zoomed-in JWST NIRCam F150W and F410M images (Figures \ref{Fig1}c and \ref{Fig1}d). We observed excellent positional alignment between the NIRCam and ALMA data for both protostars. The zoomed-in NIRCam F150W view of Ced 110 IRS4A and IRS4B (Figure \ref{Fig1}c) reveals that the ALMA continuum disk aligns with the dark lanes corresponding to the disk shadow, though the ALMA disk is slightly shifted to the northern side.  This slight shift might be because the northern side of the disk, which is the near-side, is brighter than the southern side and/or due to the opacity of dust in the midplane at 1.5 $\mu$m and 1.3mm being different.

\begin{figure*}
\centering
\includegraphics[width=0.9\linewidth]{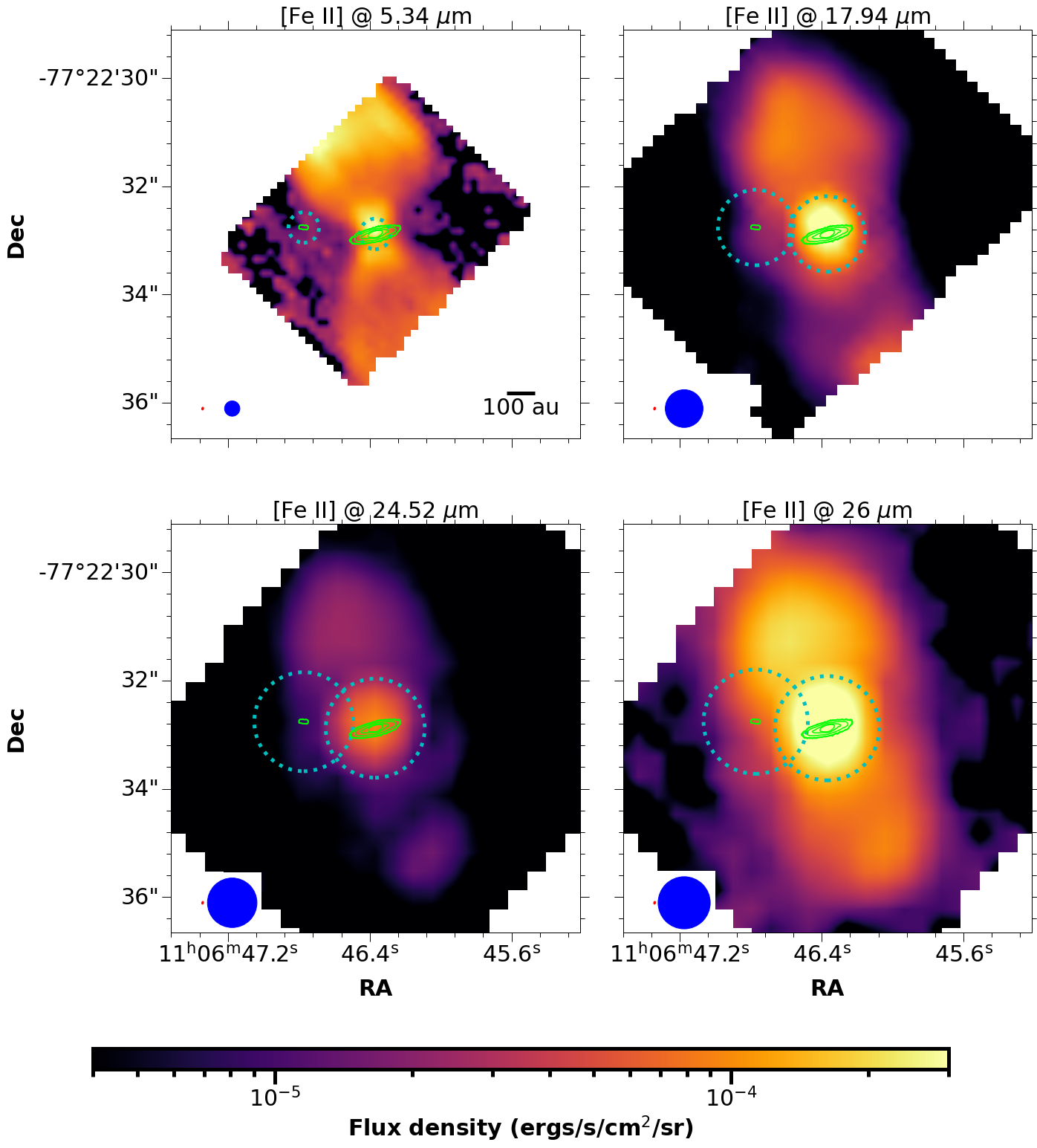}
\caption{The morphology of the [Fe II] lines detected towards Ced 110 IRS4. All images are cropped to the same spatial scale of 7.2\arcsec by 7.2\arcsec and have the same color scale. The ALMA 1.3 mm contours are shown in lime.  The scale bar corresponding to 100 au is shown in the bottom right corner. The MIRI MRS FWHM  based on \cite{2023AJ....166...45L} is shown in the bottom left corner as a blue circle and ALMA beam is shown in red. We also show in cyan apertures (radius =MIRI MRS FWHM) used to extract line flux from the two protostars. }
\label{Fig5}
\end{figure*}

In Figure \ref{Fig2a} we compare the large-scale structure seen in the JWST NIRCam F150W images with the CO isotopologues from the ALMA observations. \cite{2023ApJ...954...67S} have shown that the CO isotopologues trace both the large-scale structure and the Keplerian disk around Ced 110 IRS4. {Upon overlaying the ALMA CO (SB) isotopologue data on the NIRCam images, we observe that the large-scale structures traced are best traced in C$^{18}$O, which align well with the F150W emission.} The northern arc-like structure and the southern structures are partially traced in all CO isotopologues. The $^{13}$CO also traces the apparent bow shock at the end of the southern lobe.

In Figure \ref{Fig4}, we present a zoomed-in view of the ALMA SBLB data of CO isotopologues overlaid on the NIRCam F150W image. The gas disk, as traced by CO emissions \citep{2023ApJ...954...67S}, aligns with the disk shadow observed in the NIRCam data. This alignment underscores the powerful synergy between high-resolution ALMA and NIRCam observations, demonstrating how combining these datasets can enhance our understanding of the structure and dynamics of protostellar disks. The gas disk similar to the 1.3 mm dust also appears to be slightly shifted toward the north from the shadow.

The MIRI MRS continuum images for Ced 110 field is shown in Figure \ref{MIRI_cont}.  We made continuum integrated image at 5.38 \micron{} (by integrating from 5.35 \micron{} to 5.4 \micron{}) and at 14.14 \micron{} (by integrating from 14.1 \micron{} to  14.18 \micron{}). Ced 110 IRS4A is clearly detected in both images while  Ced 110 IRS4B is weakly detected only at 5.38 \micron{}, with strong contamination from the PSF of Ced 110 IRS4A. We however do not detect Ced 110 IRS4B at 14.14 \micron{}.

\begin{figure*}
\centering
\includegraphics[width=0.9\linewidth]{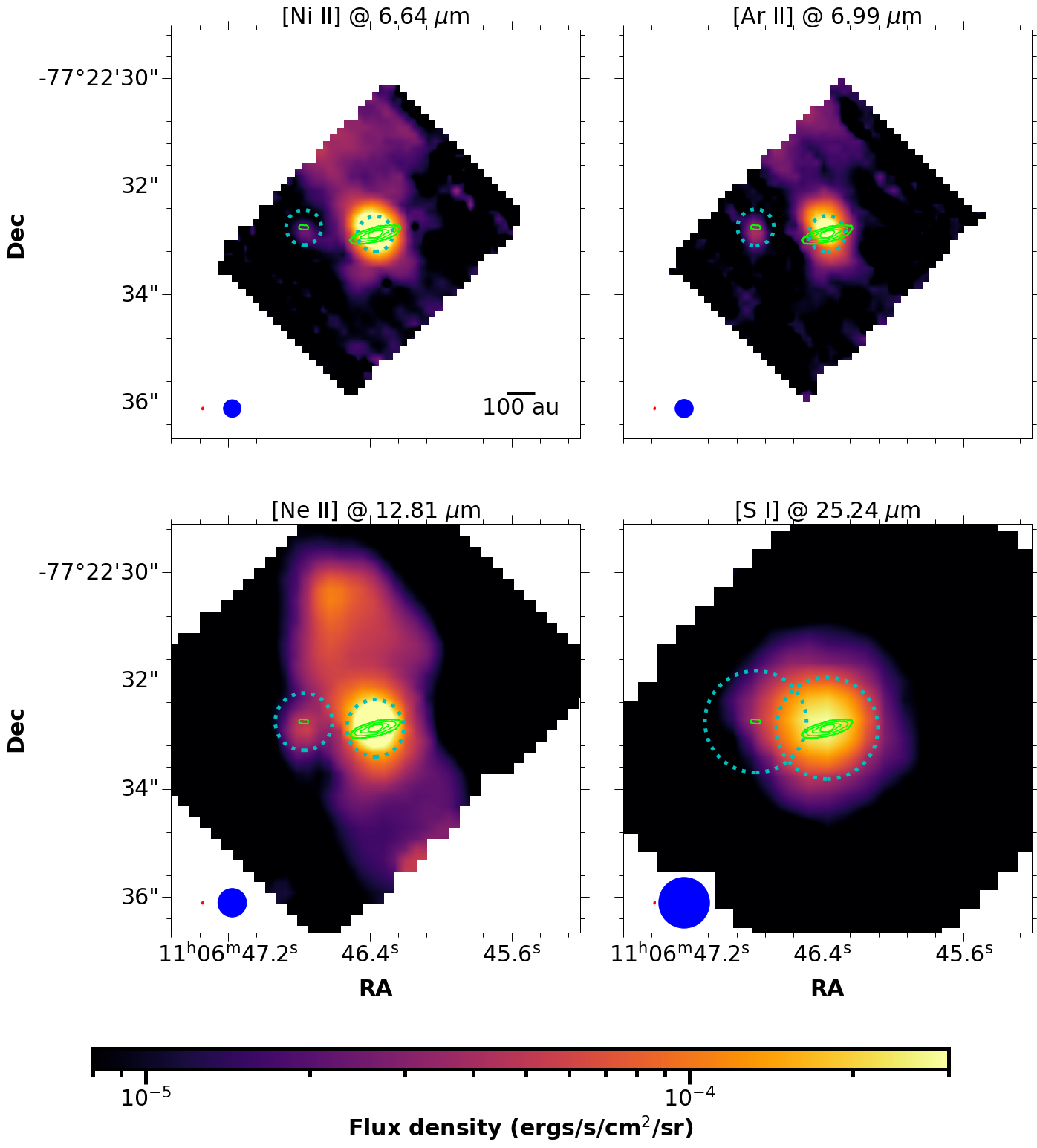}
\caption{The morphology of the atomic fine-structure lines detected towards Ced 110 IRS4. All images are cropped to the same spatial scale of 7.2\arcsec by 7.2\arcsec and have the same color scale. The ALMA 1.3 mm contours are shown in lime.  The scale bar corresponding to 100 au is shown in the bottom right corner. The MIRI MRS FWHM  based on \cite{2023AJ....166...45L} is shown in the bottom left corner as a blue circle and ALMA beam is shown in red. We also show in cyan apertures (radius =MIRI MRS FWHM) used to extract line flux from the two protostars.
}
\label{Fig6}
\end{figure*}

\subsection{Morphology of the atomic jet}

The JWST MIRI observations of Ced 110 IRS4 trace the atomic jets and molecular outflow structures down to scales of $\sim$ 50 au, comparable to the spatial resolution of ALMA observations. \cite{2023ApJ...954...67S} reported a complex emission profile around the source, with no discernible jets or outflows detected in $^{12}$CO, SO, and SiO emission. Four transitions of the [Fe II] lines were detected toward Ced 110 IRS4 (see Table \ref{TableFe}). The line maps (based on continuum-subtracted integrated line intensity) for these transitions are shown in Figure \ref{Fig5}. {The [Fe II] lines trace the atomic jet launched from the vicinity of the Ced 110 IRS4A, marking the first detection of a jet from this source. There are hints of weak [Fe II] emission from Ced 110 IRS4B especially at longer wavelengths but because of the poorer spatial resolution, it is difficult to separate the emission between Ced 110 IRS4A and Ced 110 IRS4B. In Table \ref{TableFe} we have listed the flux extracted from Ced 110 IRS4A and Ced 110 IRS4B with a radius equal to the MIRI MRS FWHM (based on \citealt{2023AJ....166...45L}) at that wavelength.  We have also overlaid the 1.3~mm continuum emission on top of the [Fe II] emission. {The jet from Ced 110 IRS4A, is perpendicular to the disk}. The [Fe II] jet (as traced in the 5.34 \micron{} line) appears to be much smaller than the disk diameter. }

The jet  expands as it moves away from the driving source, with the expansion being most evident in the [Fe II] line at 5.34 $\mu$m due to its higher angular resolution. We find that the northern part of the jet is brighter than the southern part, potentially indicating different excitation conditions or levels of extinction. \cite{2023ApJ...954...67S} also suggested that the northern side of the protostar was aligned towards us, part of this brightness asymmetry therefore can also be attributed to the geometry.

 \begin{figure*}
\centering
 \includegraphics[width=0.5\linewidth]{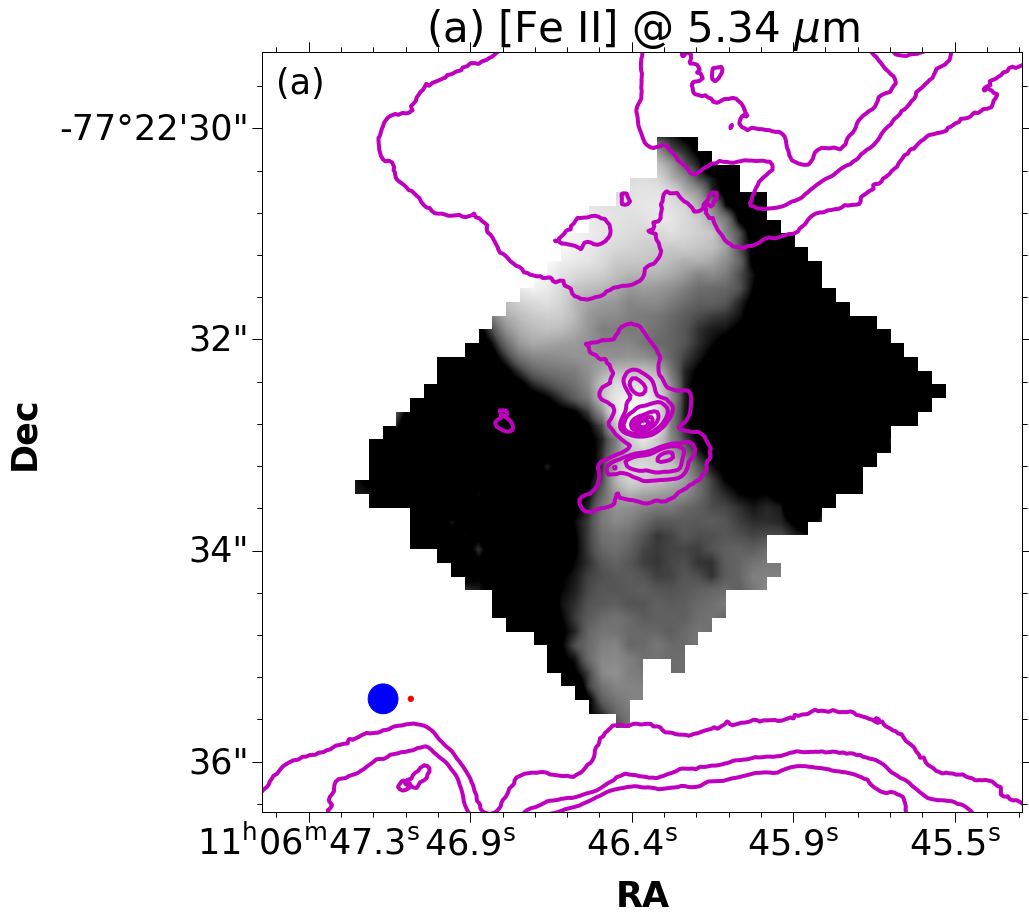}\includegraphics[width=0.5\linewidth]{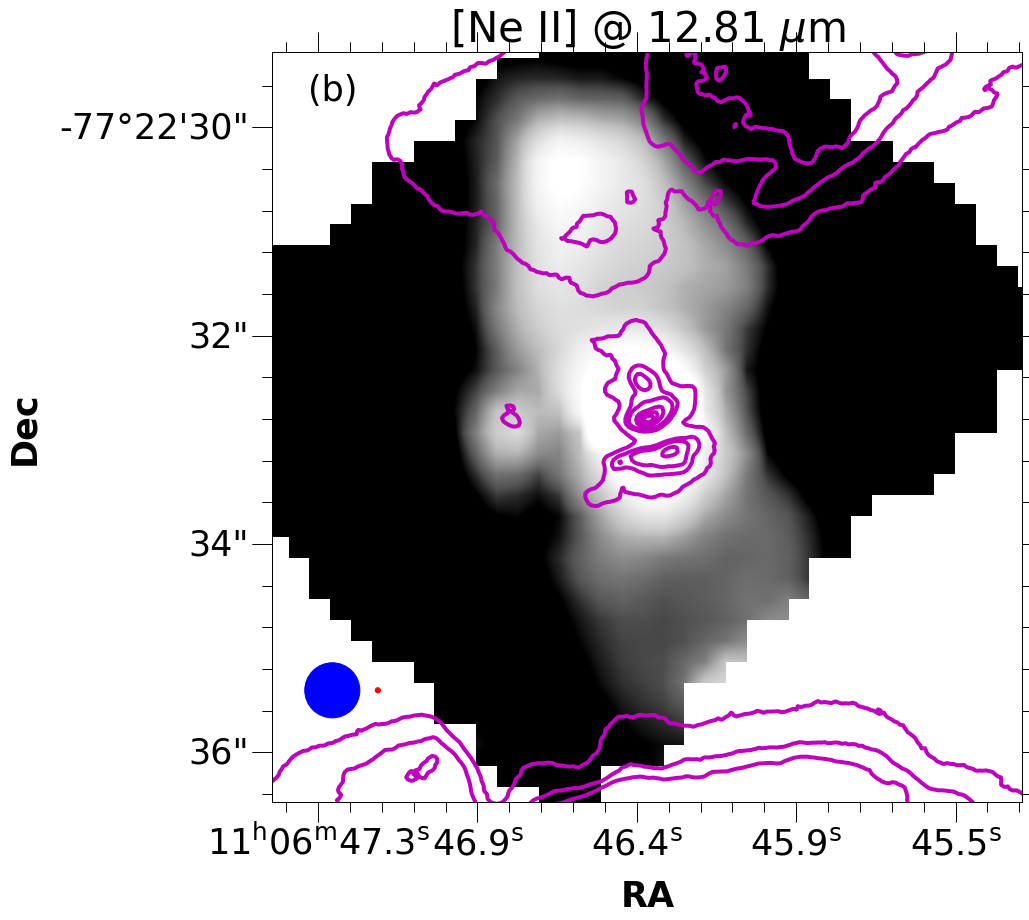}
\caption{The atomic jet has traced in [Fe II] at 5.34 $\mu$m and [Ne II] at 12.81 $\mu$m with NIRCam F150W emission overlaid on top in magenta. The MIRI FWHM based on \cite{2023AJ....166...45L} is shown in the bottom left corner as a blue circle, while the NIRCam FWHM is shown as a red circle. }
\label{jet_NIR}
\end{figure*}

In Figure \ref{Fig6}, we show the line maps of some of the fine-structure (FS) lines detected towards Ced 110 IRS4. We have listed the line flux (extarcted from apperture shown in Figure \ref{Fig6}) from the FS lines in Table. \ref{TableFe}. {The atomic and FS jet detected from Ced 110 IRS4B based on the line flux is about an order of magnitude fainter than that detected from Ced 110 IRS4A. } The jet is prominently detected in the [Ne II] line at 12.81 $\mu$m and appears to be brighter on the northern side similar to the [Fe II] jet (see Figure \ref{Fig5}). However, in other FS lines, such as [Ni II] at 6.63 $\mu$m and [Ar II] at 6.98 $\mu$m, the jet appears weaker and more concentrated around the central protostar. The [S I] emission at 25.25 $\mu$m is detected only at the central protostar. The emission detected in the [Ne II] line (Figure \ref{Fig6}) appears more extended than the disk which reflects the poorer angular resolution ( 0\farcs53 or 100 au) of MIRI at this wavelength compared to ALMA continuum observations ($0\farcs 054 \times 0\farcs 035$ or $\sim$ 10 au).  {We also find that the disk from Ced 110 IRS4B is also smaller than the emission detected in  [Ar II] at 6.99 $\mu$m, and [Ne II] at 12.81 $\mu$m. This could again be due to poorer resolution of JWST  (0\farcs28 to 0\farcs53 or 53 to 100 au) as compared to the ALMA resolution.}

We next compared the F150W NIRCam observations with the jet detected in [Fe II] at 5.34 $\mu$m and [Ne II] at 12.81 $\mu$m (see Figure \ref{jet_NIR}). We find that the emission seen in the F150W filter extends along the direction of the jet detected from the protostar, particularly towards the north. {We do not find a strong correlation between the jet, as traced by FS lines, and the large-scale structures observed in the NIRCam and ALMA data. Notably, a portion of the F150W emission near Ced 110 IRS4A overlaps with the jet detected from the protostar. However, since the MIRI observations of the jet cover only a small fraction of the NIRCam and ALMA field of view, additional observations are needed to confirm the validity of this conclusion.}

\begin{figure*}
\centering
\includegraphics[width=0.5\linewidth]{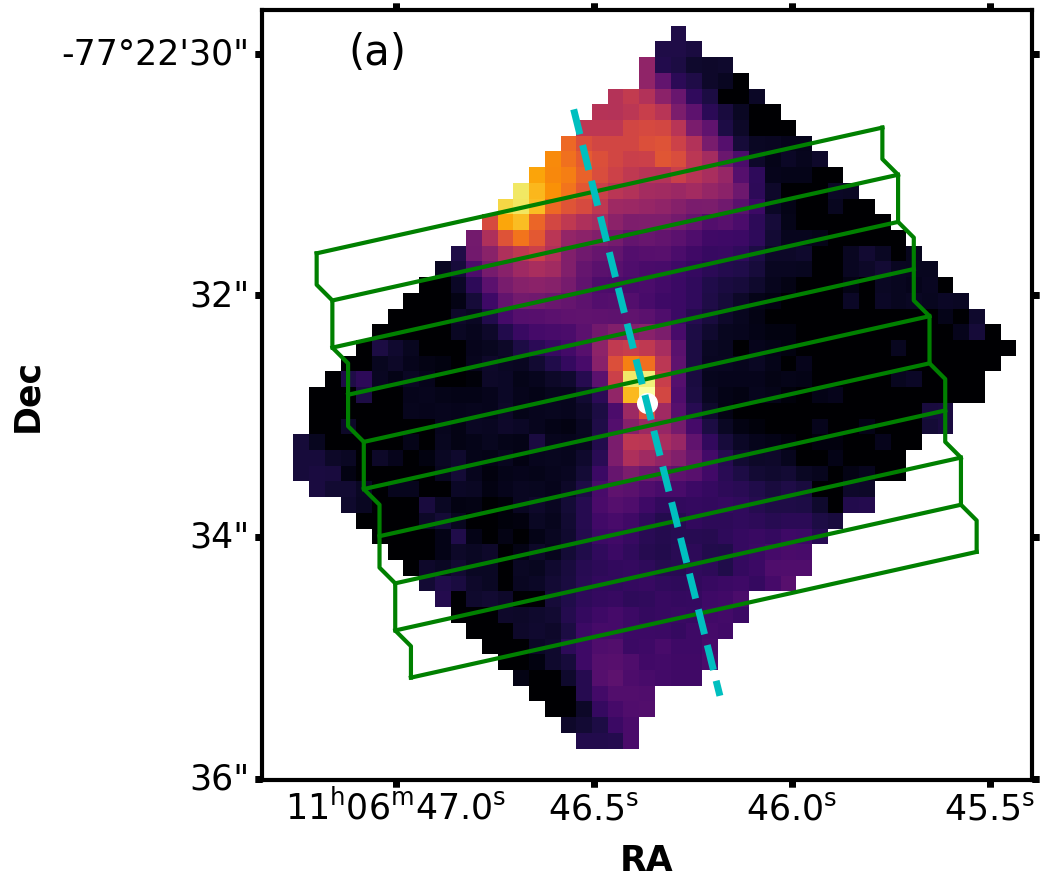}\includegraphics[width=0.5\linewidth]{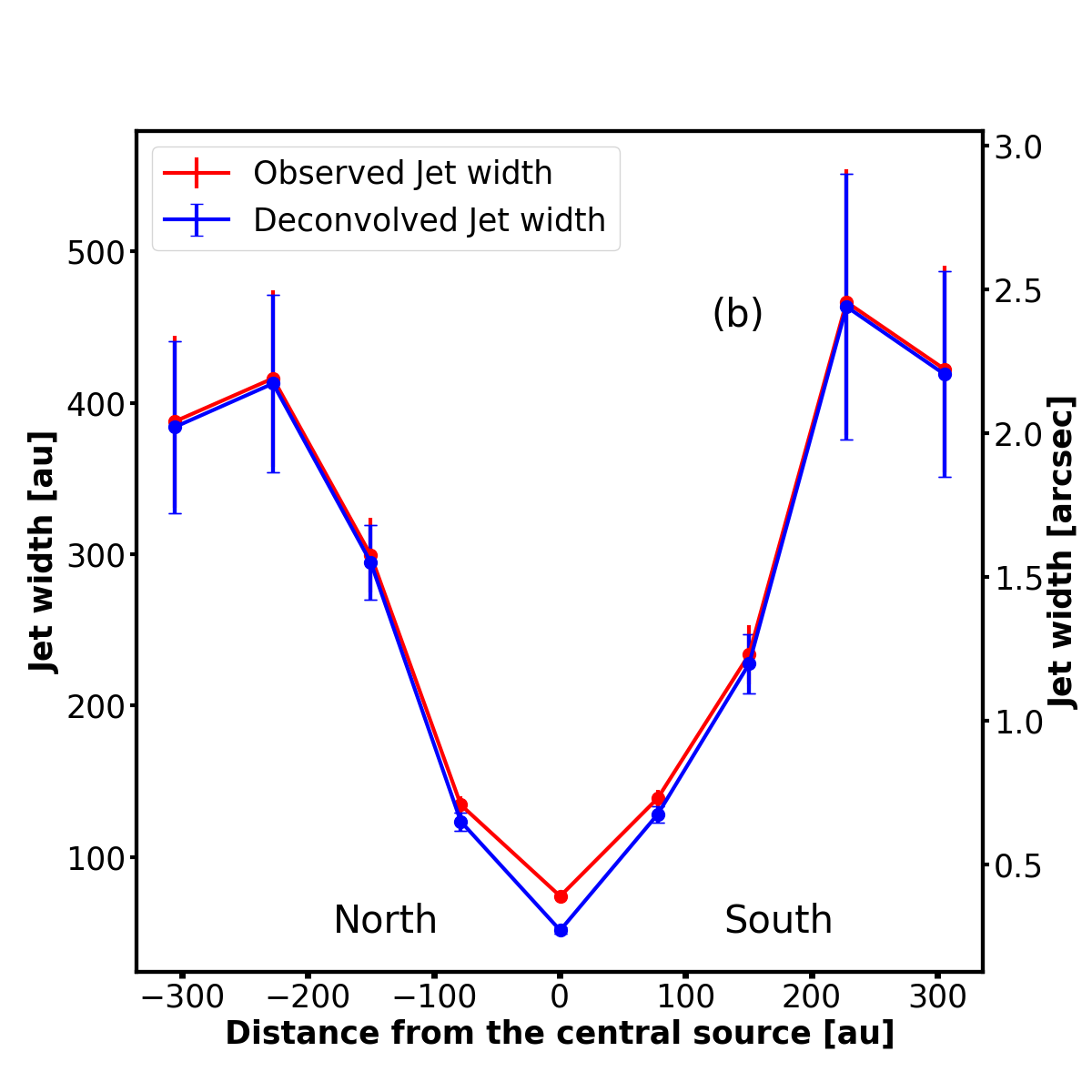}
\caption{(a) The horizontal slices in green marked on top of the jet traced in [Fe II] at 5.34~$\mu$m. The white circle represents the protostellar position. {The cyan dashed line marks the jet axis which is perpendicular to the disk as traced in ALMA dust continuum. } (b) The observed (in red) and deconvolved (in blue) jet width (FWHM) of the jet as a function of distance from the central protostar. The negative distance corresponds to the northern side, and the positive distance corresponds to the southern side.   }
\label{Fig_JW}
\end{figure*}

\begin{figure*}
\centering
 \includegraphics[width=0.8\linewidth]{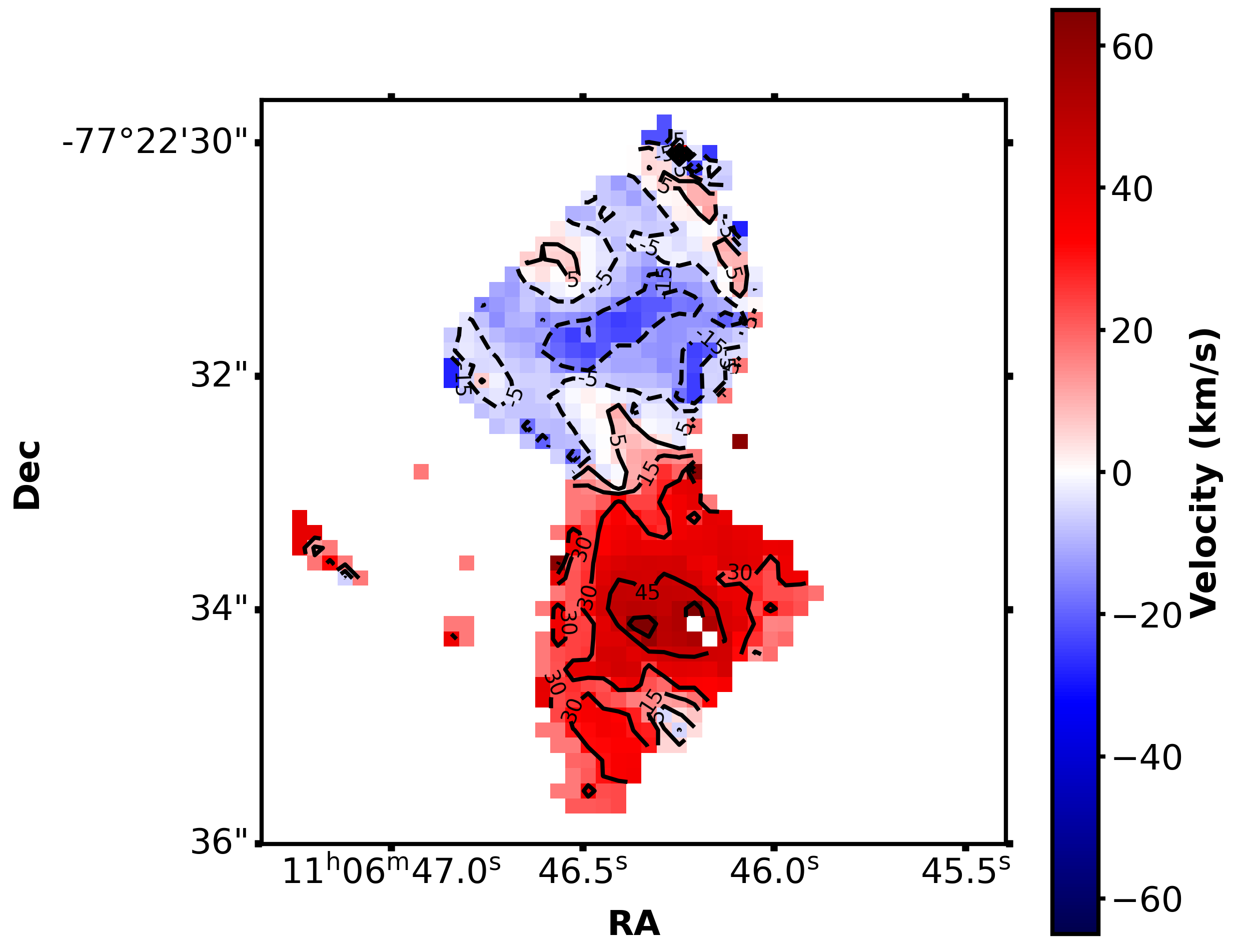}
\caption{The moment 1 map for the  [Fe II] line at 5.34~$\mu$m. {The contours mark the velocities in km\,s$^{-1}$.}}
\label{M1}
\end{figure*}

\subsection{Jet width}

At 5.34 $\mu$m the FWHM of the PSF of MIRI MRS is 0.282\arcsec \citep{2023AJ....166...45L} {or $\sim$ 53 au (at the distance of Ced 110 IRS4)}. Given the small PSF of the instrument, it is possible to measure the jet width as a function of distance from the protostar. We find that the [Fe II] at 5.34 $\mu$m  has a goblet-like morphology i.e., a long narrow neck that opens and expands rapidly as we move away from the central source. Therefore to {quantify} this expansion and determine {the opening angle of the protostellar jet, we measure the jet width as a function of distance from the protostar.}  

To measure the jet width we use slices perpendicular to the jet axis (see \citealt{2023arXiv231014061N}). Given the pixel size of 0\farcs13 and the FWHM of the PSF of MIRI MRS to be 0\farcs282, we opted to average over three pixels. These slices are shown in Figure \ref{Fig_JW}(a) and are aligned perpendicular to the jet. Only the jet's inner $\sim$400 au lies completely within the MIRI FOV.  In Figure \ref{Fig_JW}(b) we show the observed jet width as a function of distance from the protostar. The observed width of the jet is { smoothen} by the 0\arcsec.282 due to the MIRI MRS PSF. Therefore we subtract out (in quadrature) the broadening due to the PSF from the jet width to measure the deconvolved jet width. {In Figure \ref{Fig_JW}(b) we then also show the deconvolved jet width as a function of distance from the protostar. }

As shown in Figure \ref{Fig_JW}(b), the jet not only rapidly expands moving away from the center, but its opening angle, $\Theta$ (where $\Theta = \frac{1}{2}\tan^{-1} \left(\frac{\mathrm{Deconvolved\, jet\, width}}{\mathrm{Distance \, from \, the \, central \, source}}\right)$), also varies. The deconvolved jet width in the central region is $\leq51$ {$\pm$ 8.6} au. However, this measurement represents only an upper limit to the jet width due to the limited resolution of the MIRI MRS instrument. {Notably, however, this width is significantly smaller than the disk diameter of 183.4 {$\pm$ 0.4 au}, as reported by \cite{2023ApJ...954...67S}, where the dust disk radius encloses 90\% of the flux density.} Additionally, the average opening angle within the inner 150 au of the protostar (by fitting the deconvolved jet width as a function of distance) is calculated to be 23\arcdeg{} $\pm$ 4\arcdeg{}.

\begin{figure*}
\centering
 \includegraphics[width=0.5\linewidth]{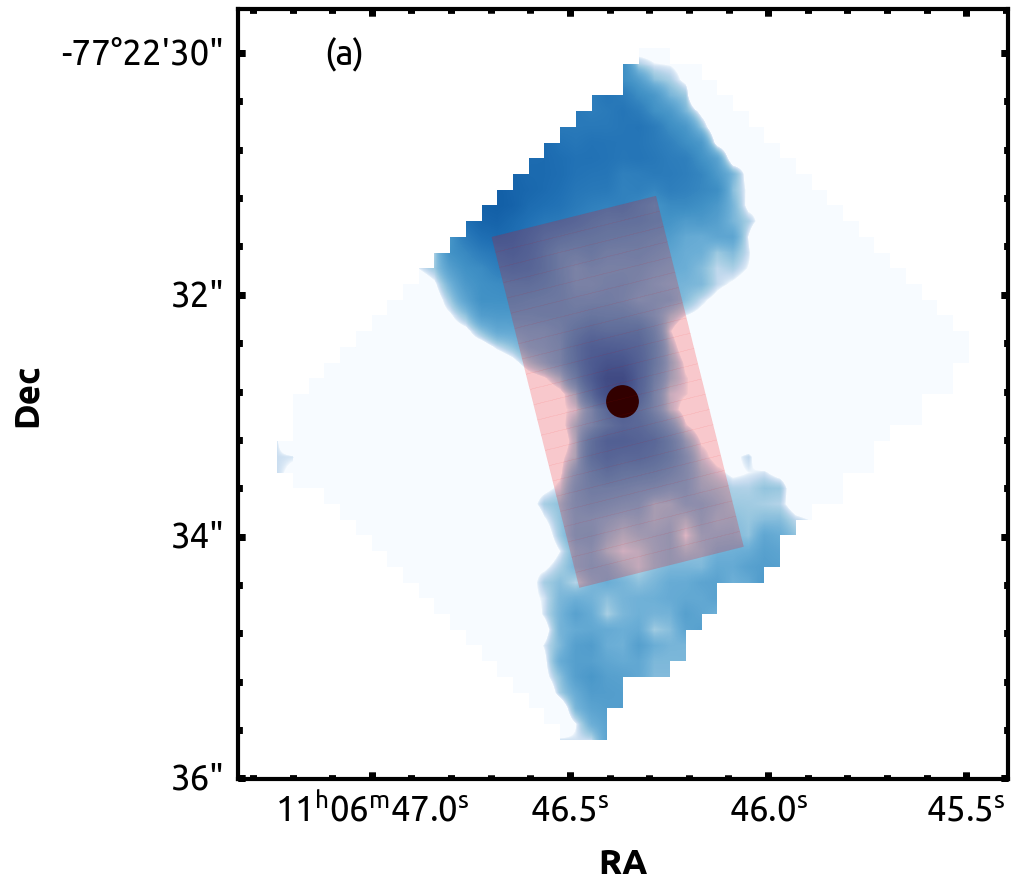}\includegraphics[width=0.5\linewidth]{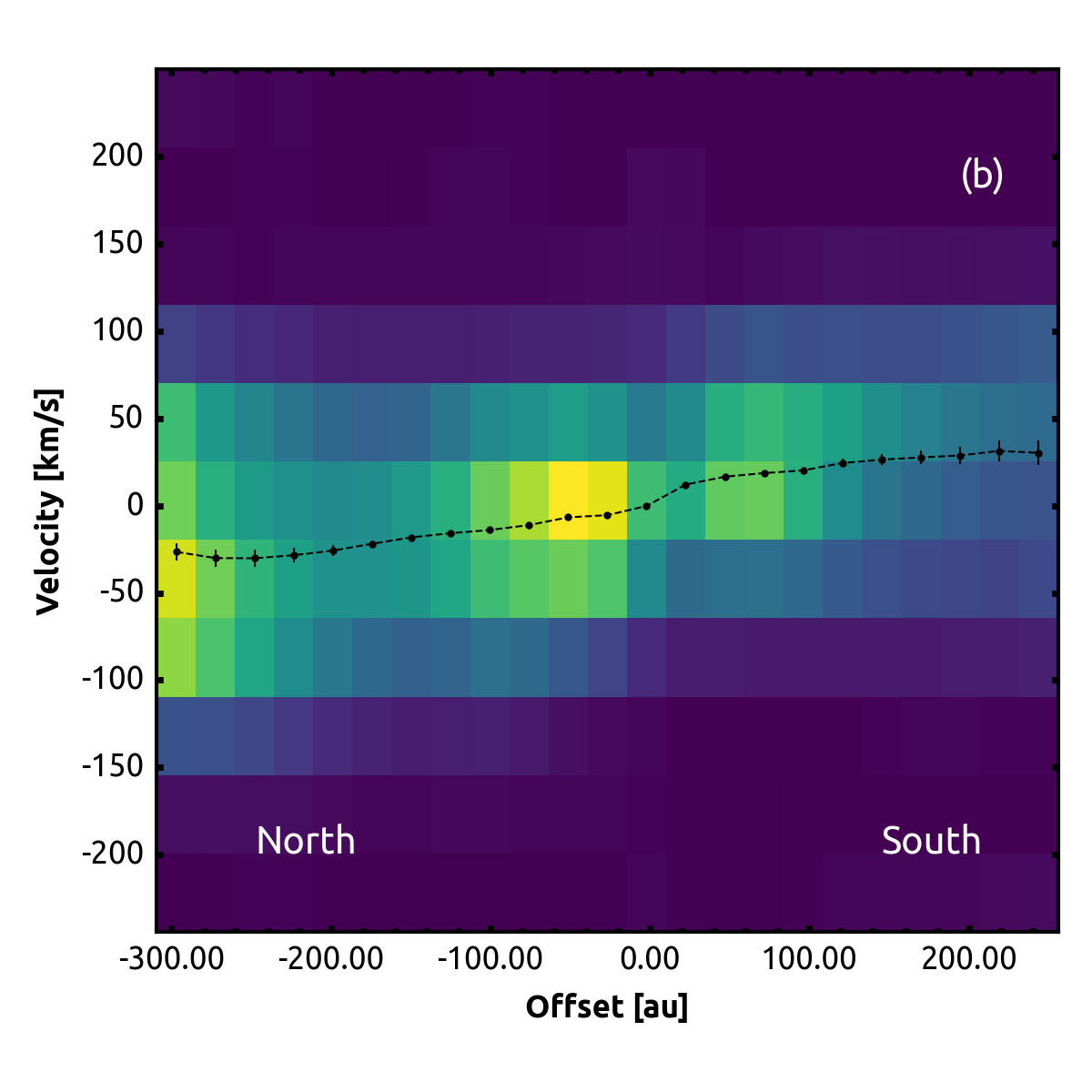}
\caption{(a) The [Fe II] jet at 5.35 $\mu$m shown in Blue overlaid with a red rectangle that denoted the region considered for making the PV diagram. The black solid circle is the ALMA continuum position. (b) The PV diagram for the [Fe II] line at 5.34 $\mu$m as a function of the distance from the central protostar along the jet axis. }
\label{PV}
\end{figure*}

\subsection{Kinematics of the atomic jet}

To explore the velocity structures within the jet, we analyzed the [Fe II] line at 5.34 $\mu$m. This line offers the best velocity as well as spatial resolution along all FS lines detected within the MIRI MRS range. In Figure \ref{M1} we show the moment 1 map of the 5.34 $\mu$m [Fe II] for Ced 110 IRS4. We find that the northern jet exhibits a blue-shifted velocity with respect to the systemic velocity (of 4.67 km\,s$^{-1}$; \citealt{2023ApJ...954...67S}); while the southern jet displays a red-shifted velocity. This means that the northern part of the protostar faces towards us similar to the conclusion from \cite{2023ApJ...954...67S}  based on the disk brightness asymmetry.  The velocity of the jet on both sides is fairly constant and the bulk blue-shift and red-shift velocity is $\sim 30-40$~km\,s$^{-1}$.  This is much larger than the systemic cloud velocity as well as the velocity of the cold gas detected with ALMA. \cite{2023ApJ...954...67S}  found that the blue-shifted and red-shifted emissions detected from molecular CO and SO both had a relatively small velocity of $\pm$ 4~km\,s$^{-1}$ with respect to the systemic cloud velocity.

To further investigate the velocity structure of the jet, we constructed a Position-Velocity (PV) diagram, for the [Fe II] line at 5.34 $\mu$m. For each slice of the spectral cube, we computed the velocity by converting the wavelength shift with respect to the lab wavelengths \citep{2016JKAS...49..109K, 2018PhRvA..98a2706T} to a velocity shift. {We meticulously chose a rectangular region that included the majority of the jet and only included the regions that were fully in the field of view, thus ensuring that no vital parts were excluded from the analysis (see Figure \ref{PV}a).} This rectangle region has a width of 1\farcs4 and a length of $\sim$3\arcsec. We further assume that the velocity of the jet is symmetric and set the average velocity of the jet to be zero. In Figure \ref{PV}(b), we show the PV diagram for the [Fe II] line at 5.34 $\mu$m. We find that the peak-to-peak velocity of the jet is $\sim$60 ~km\,s$^{-1}$. Therefore we take the velocity of the jet to be  30 ~km\,s$^{-1}$.  {The inclination angle of the protostellar disk is 76\arcdeg (based on the dust disk morphology fitting by \citet{2023ApJ...954...67S}).} If we correct the jet velocity for the inclination we find that the true velocity of the [Fe II] jet at 5.34 $\mu$m is $\sim$ 124~km\,s$^{-1}$, assuming that there is no additional radial shock component along our line of sight.

\subsection{Morphology of the molecular H$_2$ outflow}

\begin{figure*}
\centering
\includegraphics[width=0.85\linewidth]{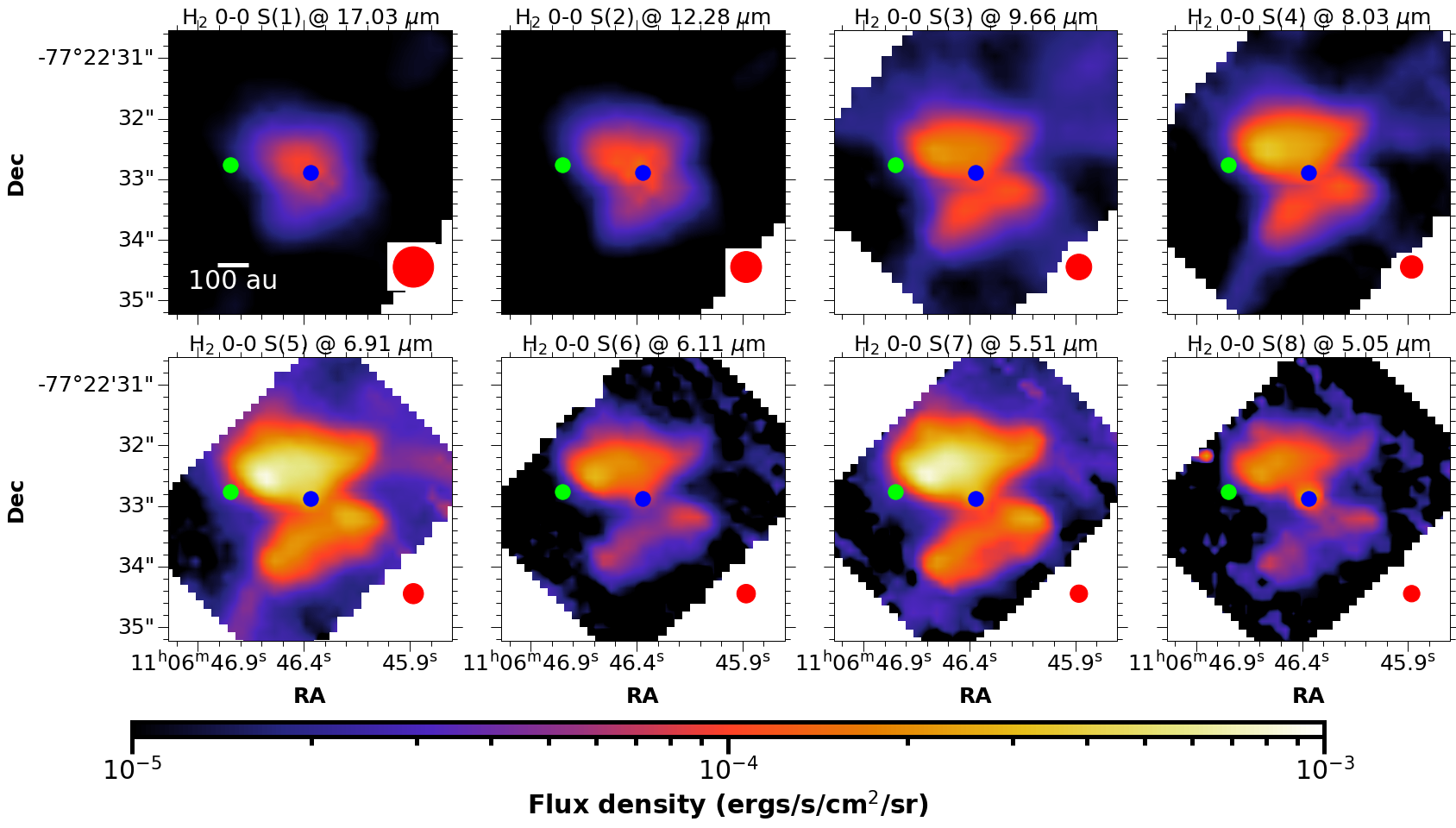}
\caption{The morphology of H$_2$  0-0 S line from S(1) to S(8) covering the MIRI MRS. All images are cropped to the same spatial scale of 5\farcs7 $\times$ 5\farcs7 and have the same color scale. The ALMA 1.3 mm position for Ced 110 IRS4A is shown as a blue circle and for IRS4B is shown as a lime circle.   The scale bar corresponding to 100 au is shown in the bottom left corner. The MIRI FWHM based on \cite{2023AJ....166...45L} is shown in the bottom right corner as a red circle.  }
\label{Fig7}
\end{figure*}

The jets and wide-angle winds launched by the disks of the protostars sweep up the infalling envelopes, entraining gas along with them.  In the swept-up gas, molecular H$_2$ gets heated in turbulent shocks. Therefore by investigating the distribution of molecular H$_2$ {emission}, we can understand the morphology of the shock-heated gas that is entrained in the protostellar outflow. We detect several pure rotational transitions of H$_2$ 0-0 S lines (see Figure \ref{Fig7}) starting from H$_2$ 0-0 S(1) to H$_2$ 0-0 S(8) (see Table \ref{TableH2}). {The molecular H2 outflow displays a distinct morphology resembling two bowls or hemispheres placed back-to-back, in stark contrast to the hourglass-like shape typically seen in protostellar outflow}s \citep[e.g.,][]{2023arXiv231003803F,2024ApJ...967..168N}. The back-to-back hemisphere-like structure gets narrower and shows a neck at the position of IRS4A. The H$_2$ emission has minima at the neck position.  In addition to the H$_2$ 0-0 S lines, the H$_2$ 1-1 S lines from S(3) to S(7) arising from the inner 2\farcs4 region around Ced-110 IRS 4 are {also} detected. {However, these lines are much weaker in strength as compared to the H$_2$ 0-0 S transitions and hence we do not analyzed these lines in this work. }

\begin{figure*}
\centering
 \includegraphics[width=0.5\linewidth]{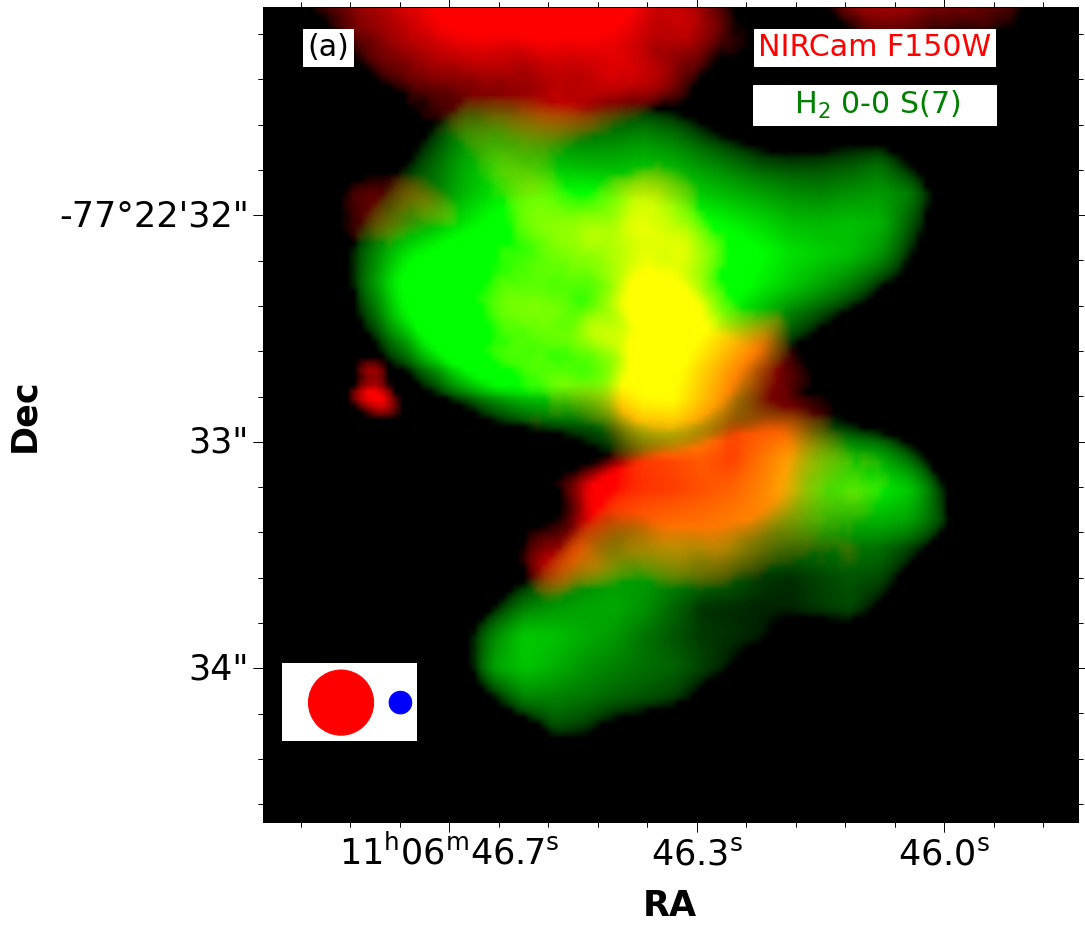}
 \includegraphics[width=0.5\linewidth]{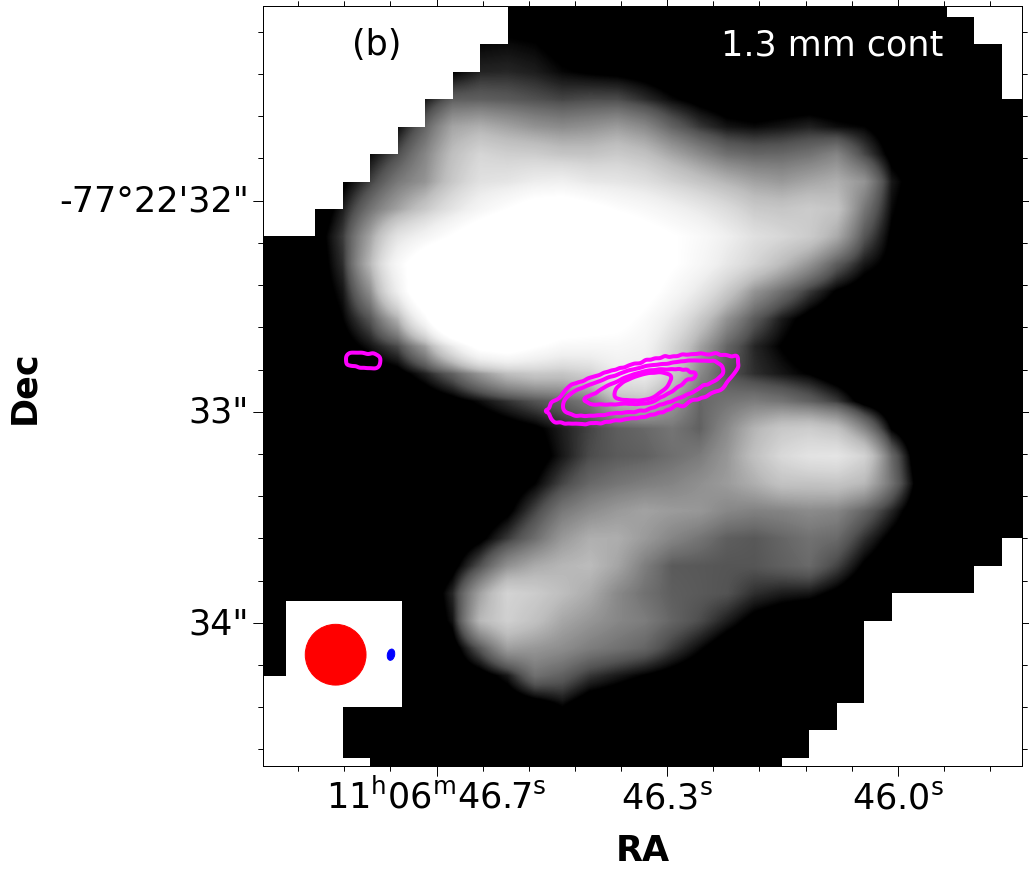}\includegraphics[width=0.5\linewidth]{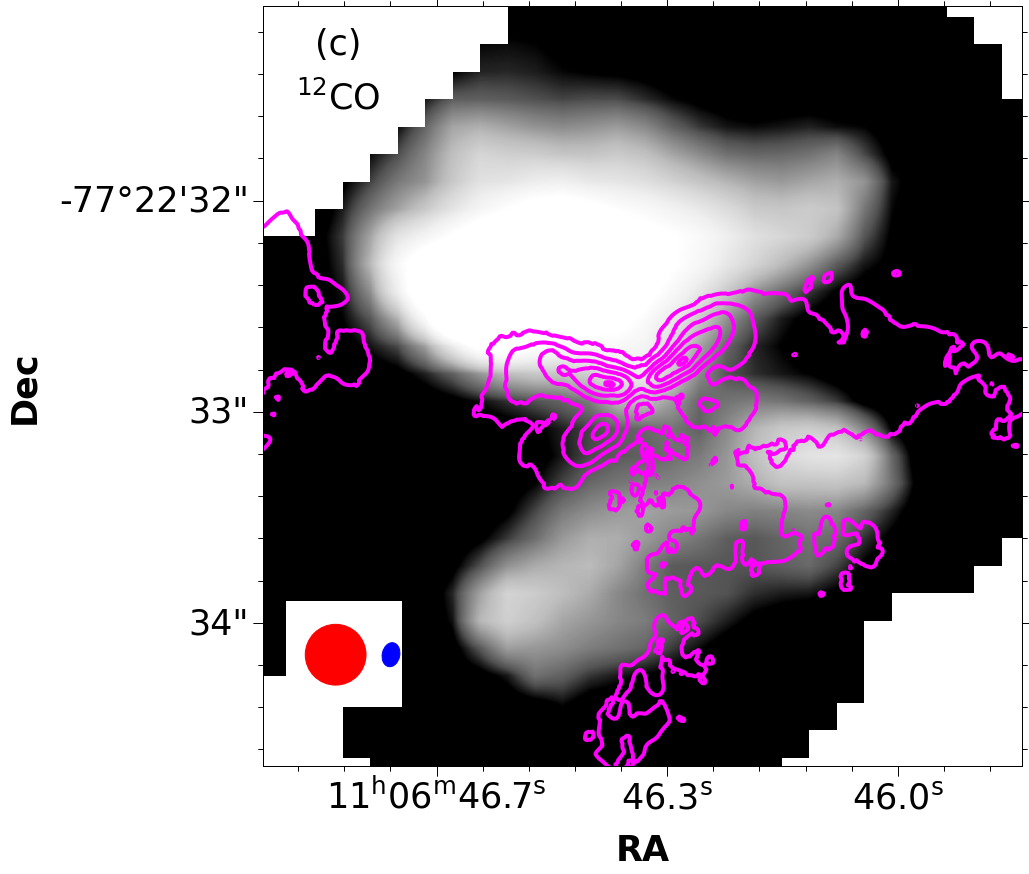}
\caption{The molecular H$_2$ 0-0 S(7) line with the NIRCam F150W emission, the SBLB 1.3 mm ALMA continuum and the ALMA $^{12}$CO observations.  The MIRI FWHM based on \cite{2023AJ....166...45L} is shown in the bottom left corner as a red circle, while the NIRCam/ALMA beam is a blue ellipse.  }
\label{Fig9}
\end{figure*}

In Figure \ref{Fig9}(a) we compared the NIRCam F150W image with the H$_2$  S(7) outflow. We find that the disk shadow as traced in NIRCam F150W aligns wells with the neck (the narrow part of the outflow that connects both hemispheres) in the F150W emission. In Figure \ref{Fig9}(b), we show the H$_2$  S(7) line with the 1.3 mm overlaid on top. {We find that the protostellar disk lies in the narrowest part of the next between  the two outflow hemispheres but slightly shifted towards the northern side.} This is consistent with the outflow geometry with the near side on the northern side. The width of the H$_2$ at the neck of the emission is comparable to the continuum disk {(see Section 4.2)}.    

{The $^{12}$CO emission appears to trace the upper layers of the disk and disk rotation \citep{2023ApJ...954...67S}.
The molecular $^{12}$CO emission from the protostar shows four intensity peaks around Ced110 IRS4A and a dark lane along the continuum major axis. The four intensity peaks likely come from a hotter disk surface on the near side and a colder disk inner surface on the {far} side of the disk \citep{2013ApJ...774...16R,2018A&A...609A..47P,2023ApJ...954...67S}.} On overplotting the $^{12}$CO on the molecular H$_2$ emission (Figure \ref{Fig9}c), we find that four intensity peaks in $^{12}$CO line up with the base of the emission of molecular H$_2$.

\section{Discussion}
\subsection{Jet from Ced 110 IRS4A}

The jet from Ced 110 IRS4A is detected in multiple transitions of [Fe II] as well as the [Ne II] line. While the jet is not completely captured in the field of view (FOV) of the [Fe II] line at 5.34 $\mu$m (in MIRI channel-1 short FOV $\sim$ 3.2\arcsec $\times$ 3.2\arcsec), the [Fe II] line at 26 $\mu$m in MIRI channel-4 (with its much larger FOV of 7.7\arcsec $\times$ 7.7\arcsec) has a spatial extent of 6.8\arcsec or 1285 au (when measured from tip to tip). If we correct for the inclination of the protostar (76\arcdeg) we find that the total jet length is 1324 au (assuming that the jet does not extend beyond the JWST footprint). From our PV analysis, we also know that the velocity of the jet is about 124~km\,s$^{-1}$ (after correcting for inclination). {Therefore we can derive a dynamical timescale of the atomic jet (withing the MIRI FOV) of $\sim25$ years. }

{Unlike the other atomic jets that have been detected with JWST from low mass stars with a typical opening angle of $\leq10\arcdeg$ (\citealt{2023arXiv231014061N}; Federman et al., in prep) the atomic jet from Ced 110 IRS4A has a much larger opening angle of $\Theta$ of 23\arcdeg{} $\pm$ 4\arcdeg{}. }This suggests that the jet is not confined. {If the jet was ballistically confined then given the jet velocity of 124 km\,s$^{-1}$ and under the assumption that the shock temperature is 10,000 K \citep[e.g.,][]{2021NewAR..9301615R} with sound speed $c_s$ = 10 km s$^{-1}$ the opening angle  $\Theta$ would be 4.6\arcdeg~, much smaller than what is measured.} \cite{2023ApJ...954...67S} found that disks of Ced 110~IRS4A and 4B are misaligned  with each other. {This misalignment may result from dynamical interactions between the primary and secondary components, potentially leading to jet precession. Such precession might, in turn, produce a  jet with a broader opening angle. This still does not explain the goblet-like structure seen in the [Fe II] line at 5.34 $\mu$m.  One possible scenario to explain the wide angle component is that in the past the launching mechanism for the [Fe II] was different than what it is today. Detailed discussion about the launching mechanism of the jet and outflow from the Ced 110 IRS4 system will be presented in a later work. }

The jet from  Ced 110 IRS4B seems to be more extended and brighter on the southern side, which is opposite to the jet from Ced 110 IRS4A. One possible explanation for this could be that the disk inclination of the IRS4B is opposite to the one of IRS4A. This suggests that the disk inclination may be different for the two protostars.

\subsection{Origin of molecular emission from Ced 110 IRS4A}

Unlike the atomic and FS emission observed from Ced 110 IRS4, the molecular H$_2$ emission from the protostar is considerably broader and lacks any collimated components.  Moreover, the distribution of H$_2$ emission around the protostar is notably asymmetric. The northern outflow region is not only brighter but also more extended compared to the {southern} region. This asymmetry indicates potential variations in the environmental conditions or the dynamics of the outflow in different directions, possibly influenced by the surrounding material.  This north-south asymmetry could also be because of the geometrical effect, i.e., the northern side of the outflow is on the near side. This geometrical configuration is consistent with the brightness asymmetry shown in the gas disk in $^{12}$CO (and probably $^{13}$CO as well) and 1.3 mm dust continuum \citep{2023ApJ...954...67S}. This configuration is also consistent with the north-south asymmetry seen in FS lines. 

In Figure \ref{Fig13}, we provide a detailed comparison of the spatial extent of H$_2$ emission across three specific rotational transitions: 0-0 S(1), 0-0 S(4), and 0-0 S(7). These transitions represent different energy levels of the H$_2$ molecule. Our analysis reveals that the extent of the outflow is similar across these transitions, suggesting a consistent outflow structure across different excitation conditions of H$_2$. {In Figure \ref{Fig13}(a) we show the contours of H$_2$ 0-0 S(4), and 0-0 S(7) overlaid on H$_2$ 0-0 S(1) the lowest lying transition of  H$_2$ available in the JWST MIRI MRS.  Interestingly, despite the higher spatial resolution of the H$_2$ 0-0 S(7) transition compared to the H$_2$ 0-0 S(1), the overall extent of the emission remains similar. }

{To quantify this result we measure the width of the H$_2$ emission across these transitions. To do so we first need to detect the edge of the outflow. For this we make use of the Sobel–Feldman operator or Sobel filter \citep{Sobel} using the implementation from \textit{`skimage'} python package  \citep{SK}. The Sobel filter is a gradient-based edge detection technique. It computes the rate of change in intensity (gradient) in both the x- and y-directions and combines them to highlight edges. It works by convolving the image with two 3x3 kernels (one for horizontal changes and one for vertical changes) and computing the gradient. We start by rotating the image using the `scipy' python package \citep{2020NatMe..17..261V} to align the outflow axis along the y-direction by rotating the image by 14\arcdeg{}.  A similar approach is also followed in \cite{2021ApJ...911..153H,HH30}. We then applied the Sobel filter on the H$_2$ images to determine the edges of the outflow.   }

{In Figure \ref{Fig13}(b) we show the implementation of the Sobel filter and the edge detection for the H$_2$ 0-0 S(4) transition. As can been seen from the Figure we are able to detect the edge of the H$_2$ emission. We next measured the width of the outflow.  Figure \ref{Fig13}(c) shows that the width of the outflow is similar across the three transitions (0-0 S(1), S(4), and S(7)) despite the varying angular resolution. This is counter to what has previously been detected in other protostars. It is found that higher J transitions of H$_2$ tend to be more collimated than the lower J transitions \citep[e.g.,  Narang et al. in prep;][]{2024A&A...687A..36T, HH30}.  This has been argued as evidence for a disk winds as the outflow launching mechanism. Thus this may suggest that MHD disk winds may not be the outflow launching mechanism. However detailed modeling of the system and its outflow are necessary to completely rule out this. }

{Using a fit to the outflow width as measured with H$_2$ 0-0 S(7) transition we can derive an upper limit to the width at the protostar.  We find that the width of the outflow at the protostar is 130 $\pm$ 10 au. This is smaller than the disk diameter of 183.4 {$\pm$ 0.4 au}, as reported by \cite{2023ApJ...954...67S}, but much larger than the upper limit of the width 51 $\pm$ 8.6 au that is derived for the [Fe II] jet.  }

\section{Summary}
In this work, we made use of NIRCam and MIRI observations from the Early Release Science Program IceAge along with millimeter continuum and gas line data from the ALMA large program eDisk to study the jet and outflow morphology from the Class 0/I protostar Ced 110 IRS4A and its companion Ced 110 IRS4B . The main results from these observations can be summarized as follows:

\begin{figure*}
\centering
 \includegraphics[width=0.5\linewidth]{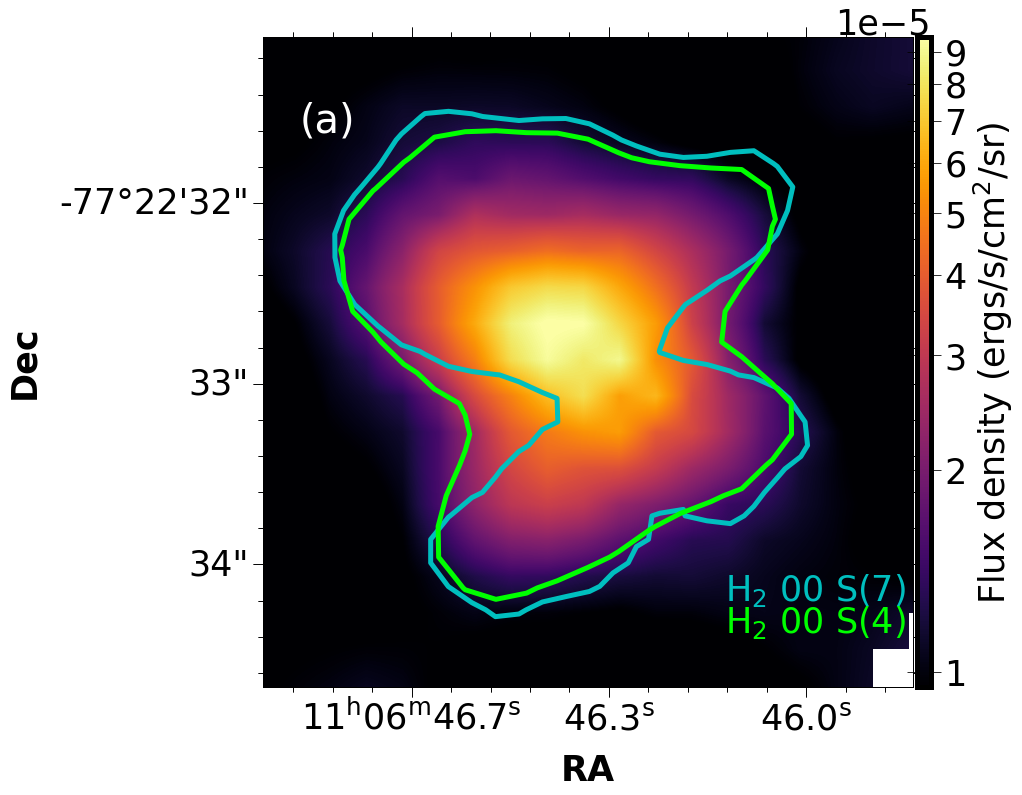}
 \includegraphics[width=0.4\linewidth]{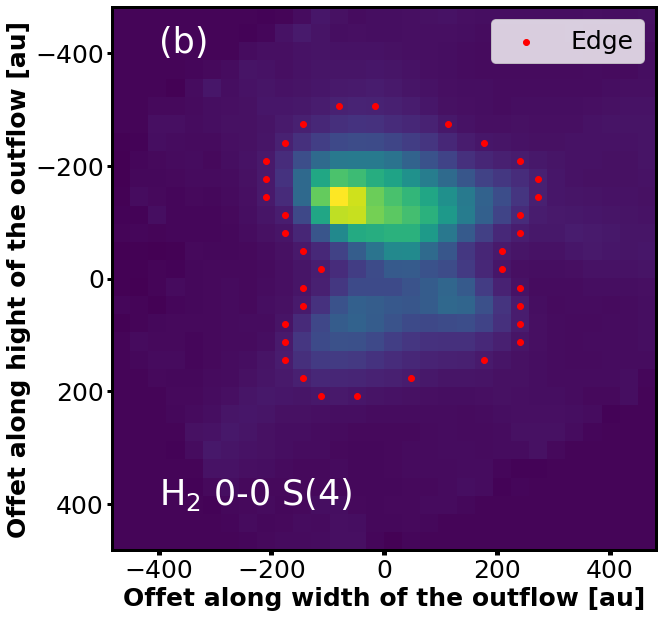}\includegraphics[width=0.4\linewidth]{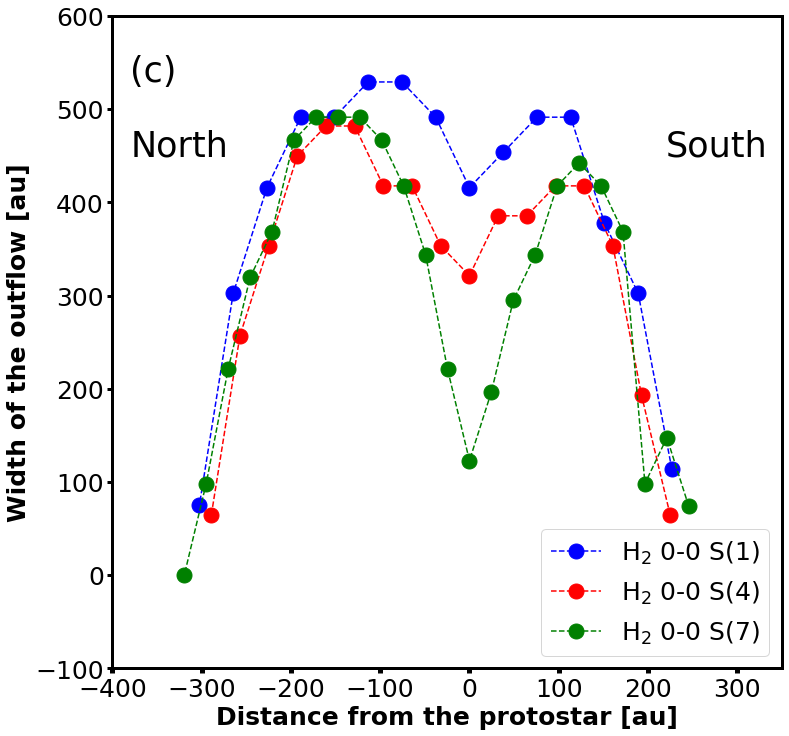}
\caption{(a) The molecular H$_2$ 0-0 S(1) line with contours of 0-0 S(4) and 0-0 S(7) overlaid on top. The  H$_2$ 0-0 S(4) and  H$_2$ 0-0 S(7) contours are at 10\% of the peak flux density (3.55 and 8.5 $\times 10^{-4}$ ergs/s/cm$^2$/sr respectively) of the emission. {(b) The edge detected (in red) on top of the H$_2$ 0-0 S(4) in color scale. (c) The outflow width as a function of distance from the protostar for the three (H$_2$ 0-0 S(1), S(4) and (7)) transitions. }  }
\label{Fig13}
\end{figure*}

\begin{enumerate}
    \item  We detect Ced 110 IRS4A and Ced 110 IRS4B  (a 0.02-0.05 $M_\odot$ substellar companion) with the NIRCam F150W and F410M observations. We also detect the disk shadow from both protostars in the F150W filter.
    
    \item Both NIRCam scattered light and ALMA observations in continuum molecular lines exhibit arc-like structures associated with Ced 110 IRS4 system.  

    \item MIRI MRS observations have detected a jet from the Class 0 protostar Ced 110 IRS4A in multiple [Fe II] lines, along with [Ni II] at 6.64 \micron{}, [Ar II] at 6.99 \micron{}, and [Ne II] at 12.81 \micron{}. This marks the first time a well-resolved jet has been detected from this system.

    \item  We also detect hints of [Fe II] as well as [Ni II] at 6.64 \micron{}, [Ar II] at 6.99 \micron{}, and [Ne II] at 12.81 \micron{} originating from Ced 110 IRS4B. This could indicate a jet from a  Ced 110 IRS4B  as well. 

    \item The jet is quite wide with the [Fe II] jet at 5.34 $\mu$m having an upper limit of the width at the base of 51 $\pm$ 8.6~au. This is much less than the disk diameter of 183.4 {$\pm$ 0.4 au}. The jet also has a large opening angle of 23\arcdeg{} $\pm$ 4\arcdeg{}, indicating a lack of collimation/confinement.

    \item  We find that the northern jet displays a blue-shifted velocity while the southern side is red-shifted. The jet has a velocity of 124 km/s, once corrected for inclination (inclination angle = 76\arcdeg; \citealt{2023ApJ...954...67S}).

    \item  The molecular H$_2$ outflow displays a distinct morphology resembling two bowls or hemispheres placed back-to-back. The emission has a minimum at the location of the protostellar disk as detected with ALMA at 1.3 mm. 

    \item The extent of molecular H$_2$ emission is similar across the different transitions, very different from what has previously been observed.  This suggests that MHD disk winds may not be the launching mechanism for the outflow detected in H$_2$.

    \item  {We find that the upper limit to the width of the outflow at the protostellar location is 130 $\pm$ 10 au which is smaller than the disk diameter of 183.4 {$\pm$ 0.4 au}  but much larger than the upper limit of the width 51 $\pm$ 8.6 au that is derived for the [Fe II] jet.}

\end{enumerate}

This work highlights the power of JWST and ALMA observations in understanding the jet and outflow processes from protostars. However further modeling of the jet and outflow is necessary to understand the wide morphology of the emission and would be taken up in a future study. 

\section{Data availability}
All of the JWST data presented in this article were obtained from the Mikulski Archive for Space Telescopes (MAST) at the Space Telescope Science Institute. The specific observations analyzed can be accessed via \dataset[DOI: 10.17909/9s60-p048]{https://doi.org/10.17909/9s60-p048}. The ALMA observations were obtained as part of the ALMA large program  Early Planet Formation in Embedded Disks eDisk (2019.1.00261.L, 2019.A.00034.S).

\facility{JWST (NIRSpec, MIRI), ALMA}

\section{Acknowledgment}
This work is based on observations made with the NASA/ESA/CSA James Webb Space Telescope. The data were obtained from the Mikulski Archive for Space Telescopes at the Space Telescope Science Institute, which is operated by the Association of Universities for Research in Astronomy, Inc., under NASA contract NAS 5-03127 for JWST. This paper makes use of the following ALMA data: ADS/ JAO.ALMA\#2019.1.00261.L;  2019.A.00034.S. ALMA is a partnership of ESO (representing its member states), NSF (USA), and NINS (Japan), together with NRC (Canada), MOST and ASIAA (Taiwan), and KASI (Republic of Korea), in cooperation with the Republic of Chile. The Joint ALMA Observatory is operated by ESO, AUI/NRAO, and NAOJ. The National Radio Astronomy Observatory is a facility of the National Science Foundation operated under cooperative agreement by Associated Universities, Inc. N.O. and M.N. acknowledge support from National Science and Technology Council (NSTC) in Taiwan (NSTC 113-2112-M-001-037). M.N. would also like to acknowledge the support of Manoj P. and Himanshu Tyagi at the Tata Institute of Fundamental Research along with the Investigating Protostellar Accretion IPA team for the detailed discussion on analysis of JWST data and reduction.

\appendix

Parameters of the ALMA maps that have been used in this analysis. 

\startlongtable 
\begin{deluxetable}{lcc}
\tablecaption{Parameters of the ALMA maps used}
\label{Table1}
\tablehead{
\colhead{Name} & \colhead{Robust} & \colhead{Beam Size (PA)}  \\
}
\startdata
\hline
\\
\multicolumn{3}{c}{Short-baseline + Long-baseline (SBLB)} \\
\\
\hline
1.3 mm continuum & 0 & $0\farcs 054 \times 0\farcs 035$ ($-12.5^\circ$)\\
1.3 mm continuum & 2 & $0\farcs 122 \times 0\farcs 086$ ($-18.9^\circ$) \\
$^{12}$CO $J=2$--1& 0.5 & $0\farcs 115 \times 0\farcs 083$ ($-12.5^\circ$)  \\
$^{13}$CO $J=2$--1 & 1 & $0\farcs 155 \times 0\farcs 107$ ($-22.2^\circ$) \\
C$^{18}$O $J=2$--1& 1 & $0\farcs 153 \times 0\farcs 107$ ($-19.4^\circ$) \\
\hline
\\
\multicolumn{3}{c}{Short-baseline (SB)} \\
\\
\hline
$^{12}$CO $J=2$--1& 0.5 & $0\farcs 465 \times 0\farcs 261$ ($13^\circ$)  \\
$^{13}$CO $J=2$--1 & 0.5 & $0\farcs 155 \times 0\farcs 107$ ($22.2^\circ$)  \\
C$^{18}$O $J=2$--1& 0.5 & $0\farcs 153 \times 0\farcs 107$ ($19.4^\circ$) \\
\enddata
\end{deluxetable}

\startlongtable 
\begin{deluxetable}{lccc}
\tablecaption{Characteristics of the NIRCam filters based on \cite{2023PASP..135d8001R} also see \href{https://jwst-docs.stsci.edu/jwst-near-infrared-camera/nircam-instrumentation/nircam-filters}{JWST NIRCam documentation} }
\label{NIRcam}
\tablehead{
\colhead{Name} & \colhead{Pivot $\lambda$} & \colhead{$\Delta \lambda$}  & \colhead{Empirical PSF FWHM}\\
\colhead{} & \colhead{(\micron)} & \colhead{(\micron)}  & \colhead{(\arcsec)}\\
}
\startdata
 F140M & 1.405 & 0.141 & 0.048 \\
 F150W & 1.501 & 0.317 & 0.050 \\
 F410M & 4.083 & 0.436 & 0.137\\
\enddata
\end{deluxetable}

The atomic and fine structure lines analyzed in this work from Ced 110 IRS4. 

\startlongtable 
\begin{deluxetable*}{ccccccccccc}
    \tablewidth{0pt}
    \tablecaption{The ionic and atomic lines reported in this work. {The FWHM of the lines based on the empirical relationship from \cite{2023AJ....166...45L}} is also provided}
    \tablehead{
    \colhead{Wavelength} & \multicolumn{4}{c}{Species and transition}  &  \colhead{$A_{ul}$} & \colhead{$E_{up}$} & \colhead{FWHM} &\colhead{Line flux IRS4A} & \colhead{Line flux IRS4B} &\colhead{ref} \\
    \colhead{($\micron$)} & \multicolumn{4}{c}{Upper state - Lower  state}  &  \colhead{(s$^{-1}$)}   &  \colhead{(K)} &  \colhead{(\arcsec)} &  \colhead{(erg $\times$ 10$^{-15}$ s$^{-1}$ cm$^{-2}$)} &  \colhead{($\times$ 10$^{-15}$ erg s$^{-1}$ cm$^{-2}$)} & \colhead{}  \\
    }
    \decimals
    \startdata
5.340	 & 	[Fe II]	 & 	${}^4 F_{9/2}-{}^6 D_{9/2}$	 & 		 & 		 & 	 2.37 $\times 10^{-5}$ 	 & 		2694 & 0.282 & 1.17 $\pm$ 0.03 & 0.07 $\pm$ 0.01	&1,2	 \\
6.636	& 	[Ni II]	 & 	${}^2 D_{3/2}-{}^2 D_{5/2}$	 & 		 & 		 & 	 5.54 $\times 10^{-2}$ 	 & 		2168 & 0.325& 7.4 $\pm$ 0.1 & 0.1 $\pm$ 0.005	& 2	 \\
6.985	 & 	[Ar II]	 & 	${}^2 P_{1/2}-{}^2 P_{3/2}$	 & 		 & 		 & 	 5.3 $\times 10^{-2}$ 	 & 		2059 &0.336	& 2.21 $\pm$ 0.02 & 0.16 $\pm$ 0.005 &2	 \\
12.814	 & 	[Ne II]	 & 	${}^2 P_{1/2}-{}^2 P_{3/2}$	 & 		 & 		 & 	 8.32 $\times 10^{-3}$ 	 & 		1123 &	0.53& 7.97 $\pm$ 0.03 & 0.72 $\pm$ 0.01& 2	 \\
17.936	 & 	[Fe II]	 & 	${}^4 F_{7/2}-{}^4 F_{9/2}$	 & 		 & 		 & 	 5.84 $\times  10^{-3}$	 & 		3496	& 0.7 & 6.88 $\pm$ 0.07 & 0.8 $\pm$ 0.01& 2		 \\
24.519	 & 	[Fe II]	 & 	${}^4 F_{5/2}-{}^4 F_{7/2}$	 & 		 & 		 & 	 3.93 $\times 10^{-3}$ 	 & 		4083 & 0.915 & 3.42 $\pm$ 0.13 & 0.81 $\pm$ 0.14&2			 \\
25.249	 & 	[S I]	 & 	${}^3 P_{1}-{}^3 P_{2}$	 & 		 & 		 & 	 1.4 $\times 10^{-3}$ 	 & 		570	&	0.94 & 12.0 $\pm$ 0.56 & 3.4 $\pm$ 0.2& 2 \\
25.988	 & 	[Fe II]	 & 	${}^6 D_{7/2}-{}^6 D_{9/2}$	 & 		 & 		 & 	 2.14 $\times 10^{-3}$ 	 & 		554	&  0.963 & 25.4 $\pm$ 1.2 & 6.5 $\pm$ 0.4& 	 1,2 \\ \enddata
    \tablecomments{(1) \cite{2018PhRvA..98a2706T}; (2)  \cite{NIST}}
     \label{TableFe}
      \vspace{-0.3in}
\end{deluxetable*}

The molecular H$_2$ lines detected in the outflow from Ced 110 IRS4. 

\startlongtable 
\begin{deluxetable*}{cccccc}
    \tablewidth{0pt}
    \tablecaption{The H$_2$  lines analyzed in this work. The wavelength, Einstein A coefficient $A_{ul}$, and upper state energy are from \cite{2022JQSRT.27707949G}. {The FWHM of the lines based on the empirical relationship from \cite{2023AJ....166...45L} is also provided. }}
    \label{TableH2}
    \tablehead{
    \colhead{Wavelength} &  \colhead{Name} &   \colhead{$A_{ul}$} & \colhead{$E_{up}$} & \colhead{FHWM} \\
    \colhead{($\micron$)} & \colhead{} &  \colhead{(s$^{-1}$)}   &  \colhead{(K)} & \colhead{(\arcsec)} \\
    }
    \decimals
    \startdata
5.053	 & 	H$_2$ 0-0 S(8)	 	 		 & 	  3.2 $\times  10^{-7}$	 &  8677 & 	0.272 \\
5.511	 & 	H$_2$ 0-0 S(7)	 	 		 & 	2.0 $\times 10^{-7}$ 	 & 		7196	& 0.288	 \\
6.109	 & 	H$_2$ 0-0 S(6)	 	 		 & 	 1.1 $\times 10^{-7}$ 	 & 		5830 & 0.307 	 \\ 
6.910	 & 	H$_2$ 0-0 S(5)	 	  		 & 	 5.9 $\times 10^{-8}$ 	 & 		4586	& 0.334	 \\
8.025	 & 	H$_2$ 0-0 S(4)           & 	 2.6 $\times 10^{-8}$ 	 & 		3474	& 0.37	 \\ 
9.665	 & 	H$_2$ 0-0 S(3)	   		 & 	 9.8 $\times 10^{-9}$ 	 & 		2503	& 0.425	 \\
12.279	 & 	H$_2$ 0-0 S(2)	 		 & 	  2.8 $\times  10^{-9}$	 & 		1681	& 0.511	 \\
17.035	 & 	H$_2$ 0-0 S(1)	 		 & 	4.8 $\times 10^{-10}$ 	 & 		1015	&0.668	
\enddata
      \vspace{-0.3in}
\end{deluxetable*}

\end{document}